\documentclass[useAMS,usenatbib]{mnras}

\usepackage{graphicx,amsmath,amssymb,hyperref}
\usepackage[T1]{fontenc}
\usepackage{aecompl}

\title[Detailed modelling of the 21-cm Forest]{Detailed modelling of the 21-cm Forest}
\author[B. Semelin]{B. Semelin$^{1}$\thanks{E-mail:
benoit.semelin@obspm.fr} \\
$^{1}$ Sorbonne Universit\'es, UPMC, LERMA, Observatoire de Paris, PSL research university, CNRS, F-75014, Paris, France}
\begin{document}

\date{2015 October 02. Received 2015 September 22; in original form 2015 July 09}

\pagerange{\pageref{firstpage}--\pageref{lastpage}} \pubyear{2002}

\maketitle

\label{firstpage}

\begin{abstract}
The $21$-cm forest is a promising probe of the Epoch of Reionization. The local state of the intergalactic medium (IGM) is encoded in the spectrum of a background source (radio-loud quasars or gamma ray burst afterglow) by absorption at the local $21$-cm wavelength, resulting in a continuous and fluctuating absorption level. Small-scale structures (filaments and minihaloes) in the IGM are responsible for the strongest absorption features. The absorption can also be modulated on large scales by inhomogeneous heating and
Wouthuysen-Field coupling. 

We present the results from a simulation that attempts to preserve the cosmological environment while resolving some of the small-scale structures (a few kpc resolution in a $50$ h$^{-1}$.Mpc box). The simulation couples the dynamics and the ionizing radiative transfer and includes X-ray and Lyman lines radiative transfer for a detailed physical modelling. As a result we find that soft X-ray self-shielding, Lyman-$\alpha$ self-shielding and shock heating all have an impact on the predicted values of the $21$-cm optical depth of moderately overdense structures like filaments. An correct treatment of the peculiar velocities is also critical. Modelling these processes seems necessary for accurate predictions and can be done only at high enough resolution. As a result, based on our fiducial model, we estimate that LOFAR should be able to detect a few (strong) absorptions features in a frequency range of a few tens of MHz for a $20$ mJy source located at z=10, while the SKA would extract a large fraction of the absorption information for the same source.

\end{abstract}


\section{Introduction}

Unveiling the Epoch of Reionization (EoR) is currently one of the most active field in observational extragalactic astronomy. The EoR is the period in the history of the universe  when the light from the first stars and galaxies progressively reionized the
cold neutral hydrogen in the Inter-Galactic Medium (IGM). Some existing and upcoming instruments focus on the sources of light (HST\footnote{\href{http://hubblesite.org}{http://hubblesite.org}}, JWST\footnote{\href{http://www.jwst.nasa.gov}{http://www.jwst.nasa.gov}}, E-ELT\footnote{\href{https://www.eso.org/public/teles-instr/e-elt}{https://www.eso.org/public/teles-instr/e-elt}}) while
others are designed to quantify the $21$-cm emission from the IGM (GMRT\footnote{\href{http://gmrt.ncra.tifr.res.in}{http://gmrt.ncra.tifr.res.in}}, PAPER\footnote{\href{http://eor.berkeley.edu}{http://eor.berkeley.edu}}, MWA\footnote{\href{http://www.mwatelescope.org}{http://www.mwatelescope.org}}, LOFAR\footnote{\href{http://www.lofar.org}{http://www.lofar.org}}, HERA\footnote{\href{http://reionization.org}{http://reionization.org}}, SKA\footnote{\href{http://www.skatelescope.org}{http://www.skatelescope.org}}). First detecting and then measuring the
statistical properties (power spectrum)  of the $21$-cm signal from the IGM in either absorption or emission against the
CMB is a challenge, mainly because of the 4 to 5 orders of magnitude brighter foregrounds. Removing these foregrounds without loosing the cosmological signal will require to measure the total signal  with high accuracy. This means an excellent instrumental calibration and a good handle on the effects of the ionosphere and on radio interferences. Imaging the signal with the SKA will
push these requirements even further. 

Another option suggested by \citet{Carilli02} is to observe the signal in absorption against a point-like bright background source such as a radio-loud Quasi-Stellar Object (QSO) or the radio afterglow of a Gamma Ray Burst (GRB).
The resulting absorption spectrum is called the $21$-cm forest.
How does observation against of point source compare with, for example, observation against the CMB during a strong absorption phase? In such a phase the spin temperature is already coupled to the kinetic temperature of the gas but the IGM is not yet much heated by X-rays. In this limiting case, the relevant signal is, in both  types of observation, proportional to the flux of the background (point-source or CMB) times the local $21$-cm optical depth of the IGM. At the typical imaging resolution with
the SKA, $5'\times 5'$ \citep{Mellema13}, the flux from the CMB at $150$ MHz is $\sim 4$ mJy per pixel, not much weaker
than the $10-20$ mJy usually considered for the (as yet hypothetical) background QSOs. However, with a spectral resolution better than $10$ kHz, $21$-cm forest observations probe modes up to $k= 10$ h.Mpc$^{-1}$, almost 2 orders of magnitude higher
than $21$-cm imaging against the CMB. The amplitude of the fluctuations in the signal may rise by a factor of 10 over those 2 orders of magnitude in $k$ \citep{Santos10, Baek10}, giving an edge to $21$-cm observations. Moreover, if the angular resolution is increased by using long base lines, the point-source signal  of the $21$-cm forest will not drop while the level of the spatially extended foregrounds will drop as the square of the angular resolution. When imaging the signal against the CMB, the signal also drops. This is possibly the biggest advantage of $21$-cm forest observations. Finally, with a much more favourable ratio between signal and foreground the calibration requirements will be less stringent.
But of course a $21$-cm forest observation samples the IGM on a single line of sight, yielding much less information than the tomographic imaging of the signal against the CMB. And while there is no uncertainty about the CMB as a background source, the existence of a population of sufficiently bright point-sources at high enough redshift to
make the $21$-cm forest detectable is an open question.     

While models predict a sufficient number of radio-loud QSOs at high redshift \citep{Haiman04,Wilman08}, the number of
currently detected objects at $z \sim 6$ does not match the predictions \citep{Banados15}. A possible explanation is
a more efficient muting of the synchrotron radio emission in QSOs toward higher redshifts as electrons loose a larger fraction of their energy to the CMB through inverse Compton scattering \citep{Afonso15, Ghisellini15}. Concerning the
radio afterglow of GRBs, \citet{Ciardi15} finds that those originating from Pop II stars would be too dim to reveal the $21$-cm forest even with the SKA, while the yet undetected GRBs from Pop III stars may be as bright as $10$ mJy at $100$ MHz and sufficient for $21$-cm forest studies. However, at most 1 every few years may occur in the fraction of the sky accessible to the SKA, namely $10^4$ square degrees \citep{Campisi11}.
 Quite obviously we cannot expect an abundance of suitable background sources and a multi-beaming
capacity for the SKA would greatly improve the chances of conducting a deep integration simultaneously with a $21$-cm tomographic survey, on a source located outside the limited  instantaneous field of view of the survey ($5^\circ\times5^\circ$)

Other than the brightness of the background source, the magnitude of the $21$-cm forest is determined by $\tau_{21}$, the optical thickness of the IGM to the $21$-cm radiation. $\tau_{21}$
depends on a number of physical constants and on the values of the neutral hydrogen number density, spin temperature and peculiar velocity gradient. Starless collapsed structures, usually called minihaloes, and milder overdensities such as filaments induce small scale fluctuations in $\tau_{21}$. Moreover, $\tau_{21}$ reaches it highest values in theses structures. Thus any attempt at modelling $\tau_{21}$ has to deal with these small scale structures while providing a
reasonable reionization environment that requires a large ($> 100$ cMpc) volume. Indeed, optical depth fluctuations on $> 100$ cMpc scale may exist in connection with X-ray heating fluctuations and Wouthuysen-Field \citep{Wouthuysen52,Field58} coupling fluctuations  \citep[see e.g.][]{Santos10, Baek10, Fialkov14b}. This wide range of relevant scales is probably why many of the exiting works on the subject use an analytical or semi-numerical modelling. The pioneering work by \citet{Carilli02}
is a notable exception, using a full numerical simulation with rather detailed physics, albeit in a small volume. Because of their small virial radius (of the order 1 ckpc) the absorption by minihaloes has been modelled using
analytical spherical profiles \citep{Furlanetto02, Furlanetto06c, Xu11, Meiksin11, Vasiliev13, Shimabukuro14}. The resulting $\tau_{21}$ values depends on the minihalo mass but the general agreement is that they should leave an
imprint  on the spectrum observable with the SKA. 

Non-virialized overdensities such as filaments depart much more 
from spherical symmetry and a numerical approach is more suited. \citet{Mack12} and \citet{Ewall-Wice14} use semi-numerical methods to study the large scale fluctuations in $\tau_{21}$. Indeed, with resolution 
respectively of $200$ ckpc and $1.5$ cMpc these works do not resolve the narrow absorption features created by filaments
or minihaloes. \citet{Xu09} and \citet{Ciardi13} use numerical simulations, running the ionizing radiative transfer as
a post-processing. \citet{Xu09} uses a prescription for the baryons rather than follow their dynamics consistently and
runs the radiative transfer at $\sim200$ ckpc resolution, while \citet{Ciardi13} computes the gas dynamics and computes
the radiative transfer at $\sim 400$ ckpc resolution in a $35$ h$^{-1}$.cMpc box and at $\sim 40$ ckpc in a $4$ h$^{-1}$.cMpc box. None of these works either couples the radiative transfer to the dynamics or self-consistently computes the Wouthuysen-Field effect on the spin temperature. Surprisingly \citet{Carilli02}, based on simulations by \citet{Gnedin00} is
the most complete in terms of physical modelling including coupled radiative hydrodynamics, molecular hydrogen formation and seemingly the computation of the Wouthuysen-Field effect (although whether it is computed homogeneously or not is not detailed).  Their simulations are however limited to $2-8$ h$^{-1}$.cMpc boxes and show surprisingly
high temperature in the neutral IGM, even at high redshift.

We will try to improve upon previous works by using a simulation in a $50$ h$^{-1}$.Mpc box, resolving most atomic cooling haloes ($1024^3$ particles) and scales down to a few ckpc \textit{both} for the dynamics and radiative transfer. The gas dynamics is fully coupled to the ionizing transfer, and the Wouthuysen-Field effect is computed in detail with a specific radiative transfer simulation in the Lyman lines as a post-processing. We will show how the detailed modelling associated with high resolution for the radiative transfer is critical in predicting the strongest absorption features occurring in the mildly overdense IGM. We do not however have the resolution to include the effect of minihaloes. 

We describe the simulation in section 2. and present our results in section 3. Section 4. is devoted to our conclusions.

\section{Numerical methods and description of the simulations}

\subsection{The LICORICE code}

The LICORICE code computes the cosmological evolution of gas, dark matter, and radiation. It is particle-based and uses
the Tree-SPH method to compute the dynamics \citep{Semelin02}. It includes star formation and feedback as subgrid recipes. The 3D radiative transfer of ionizing radiation (UV and X-rays) is implemented using the Monte-Carlo technique and is fully coupled to the dynamics. The numerical methods are detailed in \citet{Baek09} and \citet{Baek10} and the implementation has been validated in \citet{Iliev09}. Notable features of the implementation are that photons travel with the correct speed of light on an adaptive grid derived from the tree structure built to estimate the gravitational forces and that cosmological redshifting is included. This is especially important in the case of hard X-ray that can travel long distances in the IGM during the EoR before being absorbed. 

LICORICE also performs the 3D radiative transfer
for the Lyman lines series to compute the Wouthuysen-Field coupling of the spin temperature of hydrogen, a necessary step to compute the $21$-cm signal.
The implementation, also using the Monte Carlo technique, is described in \citet{Semelin07} and \citet{Vonlanthen11}.
The main difference with the modelling implemented in semi-numerical codes such as 21cmFast \citep{Mesinger11} and SimFast21 \citep{Santos10} is that the  correct line profile is used, including the extended Lorentzian wings. This is known to produce a steeper radial profile for the flux near ($< 10$ cMpc) the source \citep{Semelin07, Chuzhoy07}. But it also
allows for self-shielding effects that are absent if a Dirac-like line shape is used. We will show that such effects
are important in the case of the $21$-cm forest.

The upgrade in the version of the code used to run the present simulation is the addition of a layer of MPI parallelization on top of the previous OpenMP parallelization. Domain decomposition is performed on a Cartesian grid
without load balancing (for now). Indeed for EoR simulations in large volumes and using large MPI domains thanks to
the underlying OpenMP parallelization, fluctuation in the computation load of different MPI tasks does not
exceed $\sim 30 \%$ by the end of the simulation. For smaller computing volumes, using a larger number of domains, or
pushing the simulation toward smaller redshifts, load balancing would be needed. The main radiative hydrodynamics simulation described below was performed on $4096$ cores using $512$ MPI tasks and required approximately $300\, 000$
CPU hours.

\subsection{The radiative hydrodynamics simulation}

The main simulation has been run in a $50$ h$^{-1}$.Mpc box using $1024^3$ particles, half baryons and half dark matter. The corresponding mass for dark matter particles is $2.1\, 10^7 \, $M$_\odot$ and $4.2\, 10^6 \, $M$_\odot$ for baryonic
particles, allowing us to resolve marginally all atomic cooling haloes (i.e. haloes with masses larger than a few $10^8 \, M_\odot$). We use a standard $\Lambda$CDM cosmology with $H_0=70.4$ km.s$^{-1}$, $\Omega_M = 0.272$, $\Omega_b= 0.0455$, $\Omega_\Lambda=0.728$, $\sigma_8=0.81$, and $n_s=0.967$. The gravitational softening is $\epsilon= 1.5$ ckpc,
and we use a fixed short time step of $0.5$ Myr for the dynamics. Radiative transfer is computed with a shorter, adaptive time step as described in \citet{Baek09}. Snapshots are saved every $20$ Myr.

Star formation is allowed to occur in gas particles with an overdensity larger than 200. The formation
rate then follows a Schmidt law with exponent $1$: ${d \rho_s \over dt}= c_{\mathrm{eff}} \rho_g$, where $\rho_s$ is
the local star density, $\rho_g$ the local gas density and $c_{\mathrm{eff}}$ a tunable efficiency factor. $c_{\mathrm{eff}}$ can also be seen as the inverse of a conversion time which we set at $500$ Myr in the simulation. As a result, $0.2 \%$ of the gas has been converted into stars by the end of the simulation at $z=7.97$. 
In simulations that focus on the IGM and do not fully resolve the internal structure of galaxies, the star formation efficiency is largely degenerate with the unresolved escape fraction of ionizing photons as far as reionization history is concerned, to the point where those two quantities are often bundled  into a single photon-production-efficiency parameter in numerical works such as \citet{Iliev06}. 
Our rather low star formation efficiency is offset by the choice of a rather high escape fraction parameter ($f_{\mathrm{esc}}=0.4$, see below), yielding a reasonable reionization history. A higher star formation efficiency and lower escape fraction parameter would presumably yield a similar reionization history.

 We use a hybrid particle scheme where a fraction of a gas particle can be turned into stars at each time step. However, to avoid
linking the dynamics of gas and stars we use a gathering scheme described in \citet{Semelin02}, ensuring that whenever
the star fraction in a set of neighbouring gas particles becomes larger than a few percent, the stellar mass is gathered
in a single pure star particle. Feedback from star formation is implemented but was not activated for this simulation:
we relied on the feedback from photoheating to regulate star formation. Studying the relative efficiency of direct kinetic or thermal feedback and photo-heating is not the focus of this work: from our point of view the ionizing
photon production rate and thus the SFR are tunable parameters set to yield a reasonable ionization history.

From each particle containing a non-zero fraction of stars we propagate a number of photon packets at each time step. Even if the star fraction of a particle is $10^{-2}$, the corresponding stellar mass is still large enough to constitute
a representative stellar population. Consequently, to each source particle we assign an intrinsic luminosity, spectrum, and lifetime computed from a Salpeter Initial Mass Function (IMF) between $1.6$ and $120$ M$_\odot$ \citep[for details see][]{Baek10}. In addition each source particle emits $0.5 \%$ of its stellar luminosity in the form of X-rays
between $100$ eV and $2$ KeV with a spectral index $1.6$. This is a simple model that does no attempt to account for
the different spectra and different distributions of various possible contributors such as QSO, SNe or X-ray binaries. With our model for the stellar population, the resulting X-ray luminosity is equivalent to using $f_X=2.6$ where $f_X$ is the
X-ray production efficiency defined for example in \citet{Furlanetto06}. A constant escape fraction of $f_{\mathrm{esc}}=0.4$ is applied to ionizing UV photons and $f_{\mathrm{esc}}=1$ to X-ray photons. For each time step
we propagate a number of photon packets equal to $\mathrm{min}(2.\,10^8, 10^{5}\times \mathrm{number \, of\, source\, particles})$. In total we propagated more than $10^{11}$ ionizing photon packets in the entire simulation.  

\subsection{The Ly-$\alpha$ transfer simulation}

The 3D transfer of photons through the IGM, scattering in the Ly-$\alpha$ to Ly-$\zeta$ lines is computed using the
(currently distinct) version of LICORICE described in \citet{Semelin07} and \citet{Vonlanthen11}. Estimating the local
flux at Ly-$\alpha$ frequency everywhere in the simulation box is necessary to compute accurately the spin temperature of hydrogen and thus the $21$-cm  optical depth (see next section). Since the kinetic heating of the neutral IGM by Ly-$\alpha$ photons is weak \citep{Furlanetto06b} compared to that of X-rays for our choice of X-ray luminosity, we neglect the feedback of the Ly-$\alpha$ flux on quantities other than the spin temperature and run the Lyman lines transfer as a post-treatment of the simulation. 

For an identical number of photon packets the level of Monte Carlo noise is much higher in the case of Ly-$\alpha$ transfer than in the case of ionizing UV or X-ray transfer. Indeed, while UV and X-ray photon packets deposit a fraction
of their content in each cell along the path, a photon emitted in the Lyman band will contribute to the local Lyman-$\alpha$ flux in a single cell, where it finally reaches the Ly-$\alpha$ frequency through cosmological redshifting. Roughly speaking, for $N_p$ photons emitted in a grid with $N_c^3$ cells, each cell will be meaningfully affected by
$N_p  / N_c^2$ photons in the case of UV or X-rays and by $N_p / N_c^3$ photons in the case of Ly-$\alpha$. In previous works \citep{Baek10, Vonlanthen11, Zawada14} we could content ourselves with a high level of noise in each cell since a pixel of a $21$-cm brightness temperature map computed even at the SKA resolution (a few cMpc) would average over a number of simulation cells. In the case of the $21$-cm forest, the absorption features are to be observed on much smaller scale so we need to reduce the Monte Carlo noise as much as possible. Consequently we performed the Ly-$\alpha$ on a uniform $512^3$ grid, emitting 25 billion photons packets between each snapshot (for a total of more than 400 billion photon packets). This yields an average noise level smaller than $10 \%$ for the
Ly-$\alpha$ flux at a resolution of $\sim 100$ h$^{-1}$.ckpc. As we will see, this unprecedented effort allowed us to reveal self-shielding
effects in the structures responsible for the strong $21$-cm absorption features. This simulation required close to $200\, 000$ CPU hours.

\subsection{Computing the $21$-cm optical depth }

The $21$-cm forest signal is determined by two independent quantities: the continuum level of the background source and the optical depth of the IGM in the local $21$-cm line. This work focuses on the latter while the former is mostly
unknown at these high redshifts and is usually extrapolated from information at lower redshifts. The $21$-cm
optical depth of a patch of neutral hydrogen located at redshift $z$ along the line of sight can be written \citep[e.g.][]{Madau97,Furlanetto06}:

\begin{eqnarray}
\tau_{21}\!\!\!\!&= &\!\!\!{3 \over 32 \pi}{h c^3 A_{10} \over k_B \nu_0} {n_{\mathrm{HI}} \over H(z) T_s} \left( 1 + {1 \over H(z)}{dv_{\parallel} \over dl}\right)^{-1}\\
&\simeq & \!\!\!0.0092\, (1+z)^{3 \over 2} (1-x_i)\, {1+\delta \over T_s} \left( 1 + {1 \over H(z)}{dv_{\parallel} \over dl}\right)^{-1}
\end{eqnarray}
where $\nu_0= 1420.4$ MHz is the rest frame line frequency, $A_{10}= 2.85\, 10^{-15}$ s$^{-1}$ the Einstein coefficient of the corresponding hyperfine transition, $H(z)$ is the Hubble parameter, $n_H$ the local neutral hydrogen number density, $T_s$ the local hydrogen spin temperature, $\delta$ the local baryon overdensity, and $x_i$ the local hydrogen ionization fraction. Finally, ${dv_{\parallel} \over dl}$ is the peculiar velocity gradient along the line of sight, with $dv_{\parallel}$ and $dl$ both proper or both comoving. The numerical factor in the second line of the formula is evaluated for the cosmology used in the simulation. The spin
temperature can be evaluated using:

\begin{equation}
T_s^{-1}= { T_{\mathrm{CMB}}^{-1}(z) + x_\alpha T_c^{-1} + x_c T_K^{-1} \over 1 + x_\alpha + x_c }
\end{equation} 
where $T_{\mathrm{CMB}}(z)$ is the temperature of the CMB at redshift $z$, $x_c$ is the collisional coupling coefficient \citep[we use the fitting formula in][]{Kuhlen06}, $T_K$ is the kinetic temperature of the gas, $x_\alpha$ is the coupling coefficient associated with Lyman-$\alpha$ pumping and proportional to the local Lyman-$\alpha$ flux $J_\alpha$, and $T_c$ is the colour temperature of the spectrum around the Lyman-$\alpha$ line. We use the procedure presented in \citet{Hirata06} for computing $x_\alpha$ and $T_c$ from $J_\alpha$, $\delta$, $x_i$, and $T_K$. 

In this work we will present estimations of $\tau_{21}$ produced with different levels of detail and accuracy. Namely, we will either use the actual value of $T_s$ or approximate it to $T_s=T_K$ and we will include the effect of peculiar velocity gradients or not. The formulation of the effect of velocity gradients in eq. (1) is the result of a specific choice of integration variable along the line of sight. While it is a common formulation it has the
drawback of producing an unphysical divergence for negative velocity gradients equal to the Hubble flow \citep{Mao12}.
To avoid this issue we use a scheme very similar to the PPM-RRM scheme presented in \citet{Mao12}. The essence of the scheme
is to move the gas particle by a comoving distance of ${1+z \over H(z)}v_{\parallel}$ along the line of sight and recompute the density field in the resulting so-called redshift-space. This density can then be used to compute the
correct optical depth without the need to include an approximate term to account for velocities. 

\begin{table}
\begin{tabular}{| c | c | c |}
\hline
Models & $T_s$ computation & Peculiar velocities included \\
\hline
Model 1 & $T_s=T_K$ & no \\
\hline
Model 2 & $T_s(x_\alpha,\delta,x_i,T_K)$ & no \\
\hline
Model 3 & $T_s=T_K$ & yes \\
\hline
Model 4 & $T_s(x_\alpha,\delta,x_i,T_K)$ & yes \\
\hline
\end{tabular}
\caption{Different levels of modelling used in this work for computing $\tau_{21}$ as defined by Eq. (1). }
\label{model_table}
\end{table}

Note that both the natural and the thermal line width of the $21$-cm absorption line are mapped in redshift-space to $\delta z$ much smaller than our cell size. Consequently, quantities from a single cell are used to compute the local value of $\tau_{21}$.
For each snapshot of the simulation, $\tau_{21}$ was computed on a $2048^3$ grid, an intermediate choice between the
higher spatial resolution that a $1024^3$ particle-based simulation can produce and the $512^3$ resolution of the
$x_\alpha$ data. We did this for the 4 types of modelling described in table \ref{model_table}.

\section{Results}

We will first present the global history of several quantities computed from the simulation to allow a baseline comparison with other works and then highlight the effect of our detailed physical modelling on the predictions of the $21$-cm
optical depth. 
\subsection{The global history}

\begin{figure}
\includegraphics[width=8cm]{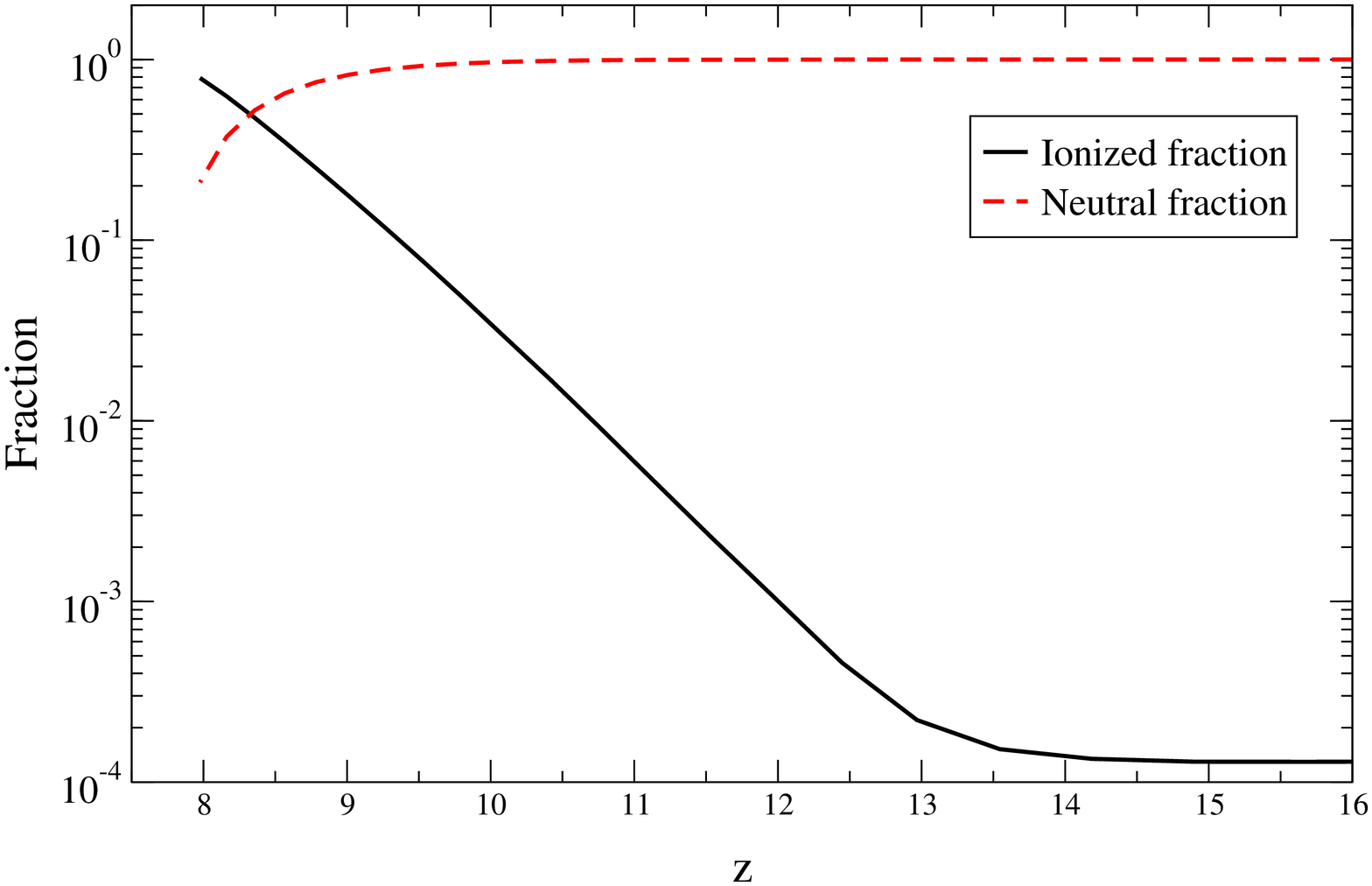}

\includegraphics[width=8cm]{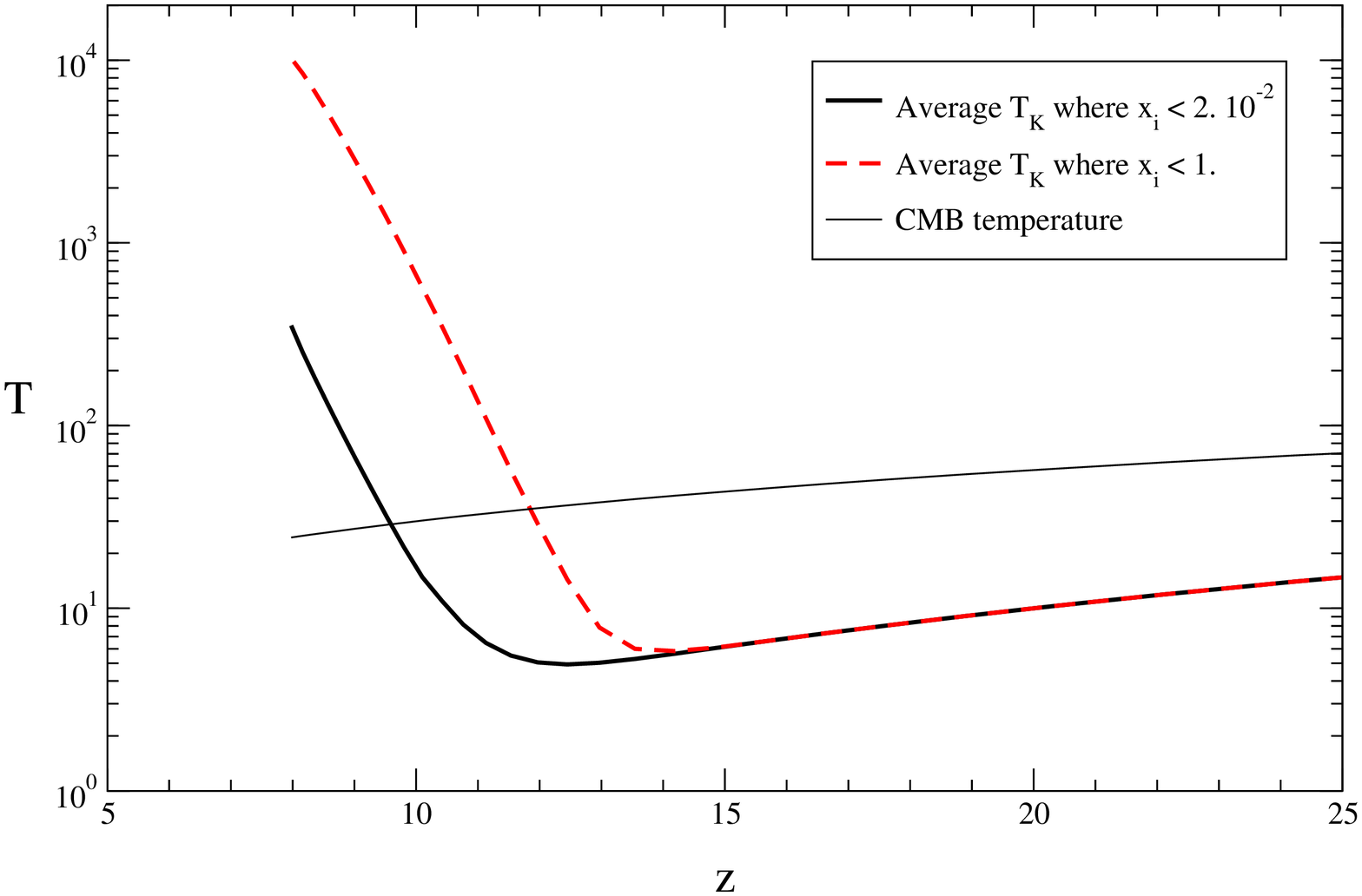}
\caption{Upper panel: history of the volume averaged ionized and neutral gas fractions. Lower panel: evolution of the average temperature of both neutral gas ($x_i < 0.02$) and neutral $+$ ionized gas. }
\label{xi_hist}
\end{figure}

\begin{figure}
\includegraphics[width=8cm]{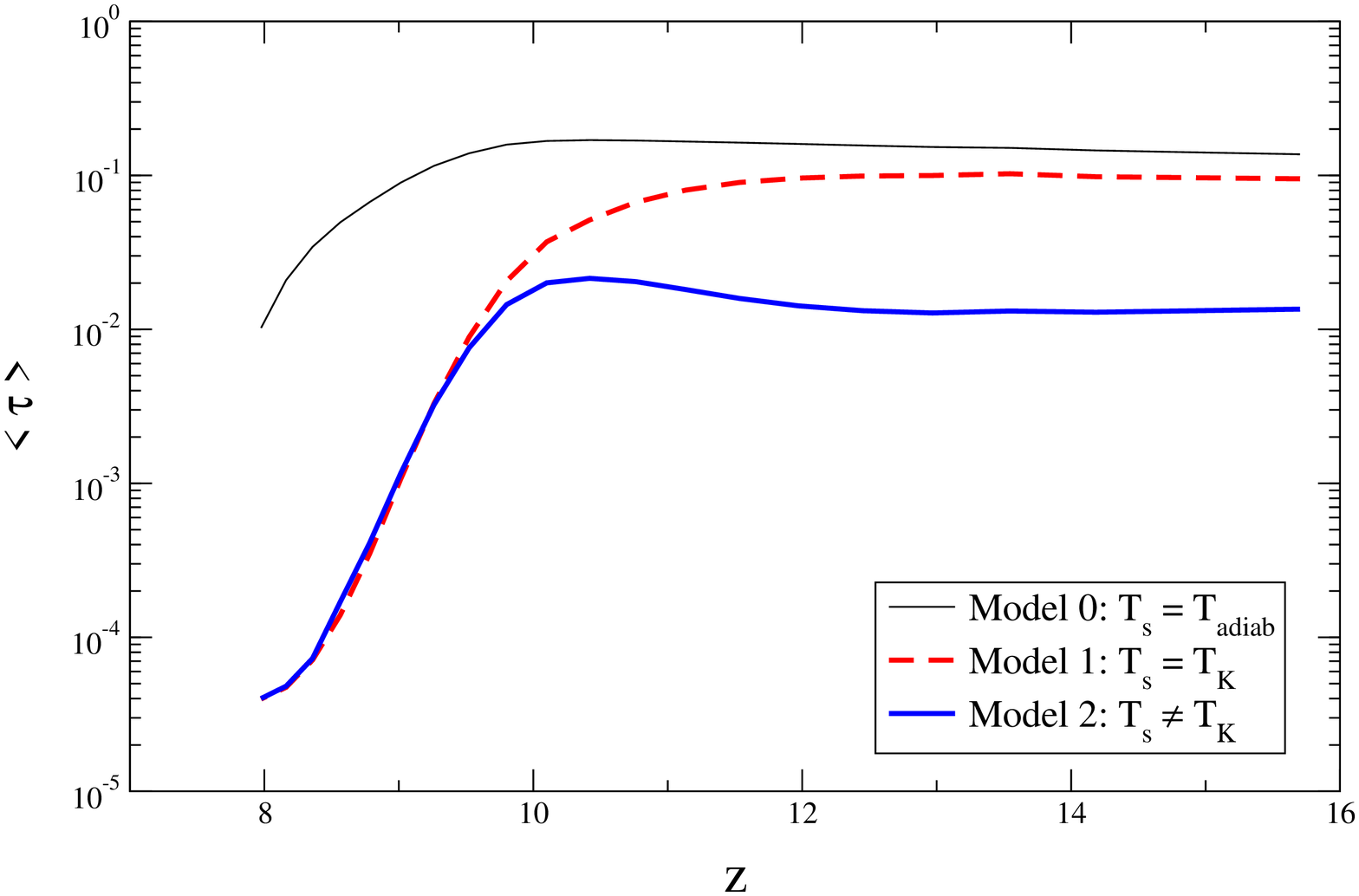}

\includegraphics[width=8cm]{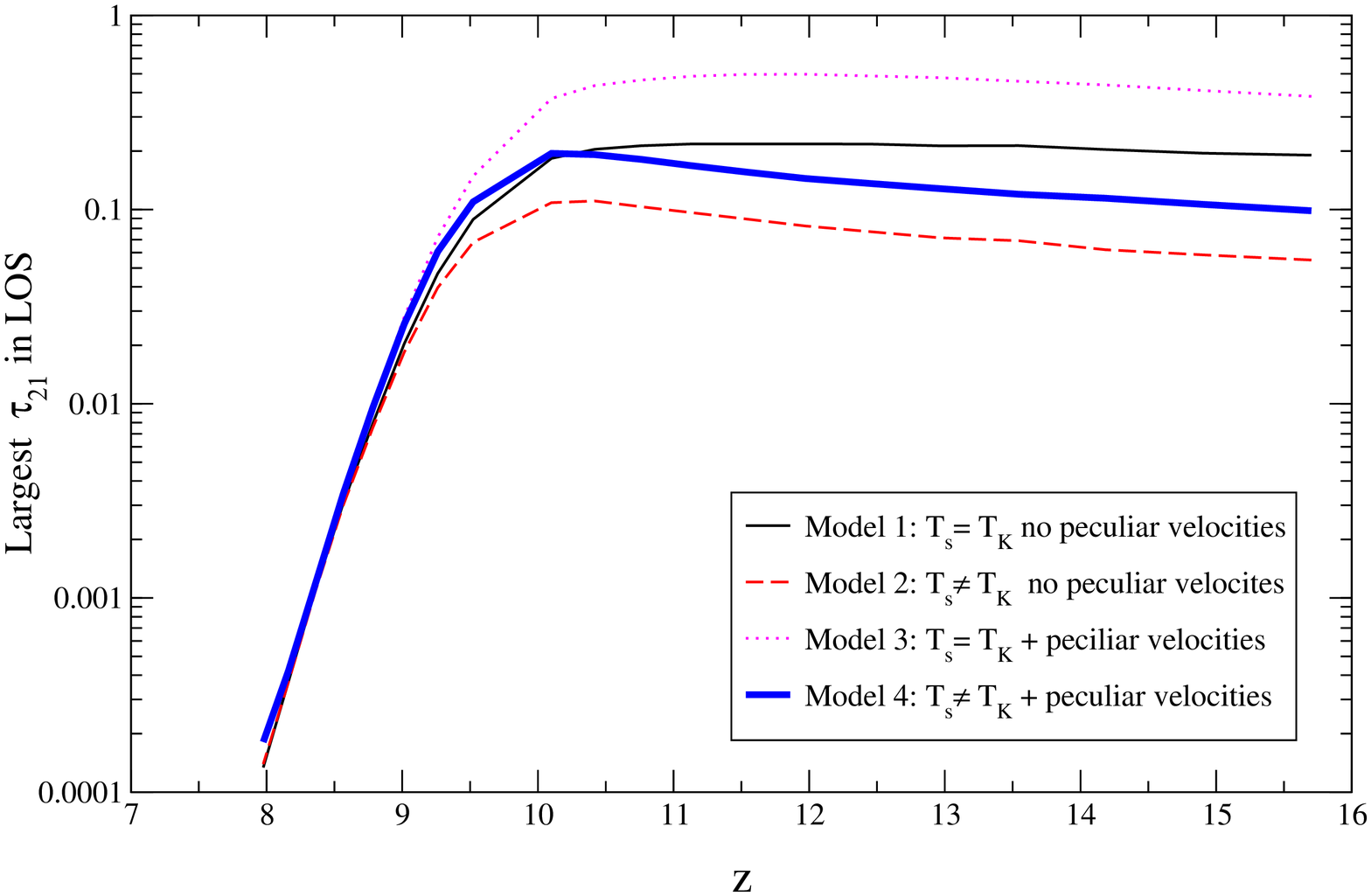}
\caption{Upper panel: evolution of the box-averaged $\tau_{21}$ for different levels of modelling as a function of redshift. Lower panel: evolution of the typical strongest absorption features in a given line of sight (see main text for definition) as a function of redshift for different levels of modelling.}
\label{average_tau}
\end{figure}

\subsubsection*{Ionization fraction and temperature}
The simulation has been run down to $z=7.97$ when the box is $80\%$ ionized (by then, X-ray heating of the neutral IGM is such that the $21$-cm optical depth is negligible). The history of the volume averaged ionized and neutral fractions of
hydrogen is plotted in the upper panel of Fig. \ref{xi_hist}. This may be considered an early reionization history.
Indeed observations of the Gunn-Peterson absorption trough in QSO spectra point to an end of reionization at $z\sim 6$ \citep{Fan06}. It should be noted that the interpretation of these observations is somewhat uncertain. They probe only the very end of reionization (neutral fraction $ < 10^{-3}$) on a small sample of biased lines of sight. \citet{Mesinger10} argue for a possible later end to reionization. The Planck 2015 value for the Thompson scattering optical depth $\tau=0.066\pm0.016$ \citep{Planck15} indicates a later reionization than previous WMAP results and decrease the tension with QSO spectra constraints. Assuming complete ionization after our last snapshot, our simulation (performed before the Planck 2015 release and consistent with the previous value of $\tau$) yields $\tau=0.078$, pointing again at a moderately early reionization history. However shifting the whole history by $\Delta z \sim 1$ would induce variations no larger than 10-20 $\%$ in the computed $21$-cm optical depth values. 

The lower panel of Fig. \ref{xi_hist} shows the average temperature history of the neutral IGM (defined as having an ionization fraction $x_i < 0.02$). At such low temperature radiative cooling is negligible and the gas initially evolves adiabatically with the expansion of the universe and the formation of structures. After the first sources are formed in the simulation box ($z \sim 15$), X-ray
heating kicks in and is the dominant contribution to the gas temperature evolution at $z < 10$. By the end of the simulation the
average temperature in the neutral regions is larger than $100 $K, very much annihilating  the prospects for detecting
the $21$-cm forest. Conservatively, we chose a moderately strong X-ray production efficiency. With a lower value, heating would be delayed and $21$-cm optical depth values would be larger.  

\subsubsection*{$21$-cm optical depth} 

The upper panel of Fig. \ref{average_tau} shows the evolution of the volume averaged value of $\tau_{21}$ for three
different assumptions for computing $T_s$. Model $1$ assumes that $T_s=T_K$ (equivalent to $x_\alpha \gg 1)$ and thus
ignores results from the Ly-$\alpha$ transfer simulation (an approximation usually valid for $\langle x_i \rangle > 0.1$), while model $2$ includes the complete computation of $T_s$. Model $0$ assumes $x_\alpha \gg 1$ and that the gas
was never heated by X-rays or shocks and thus followed an adiabatic evolution. In some sense model 0 shows the maximum possible value for $\langle \tau_{21}\rangle (z)$. When including peculiar velocities, the effect on regions in expansion (voids) and regions in contraction (density peaks) averages out and we see no difference to the curves {\sl for the average value of} $\tau_{21}$. 

Obviously, including the correct computation of $T_s$ yields
a value close to $T_{\mathrm{CMB}}$ early on, higher than $T_K$, and thus leads to a weaker average absorption in the $21$-cm forest.
In our simulation the maximum value for $\langle \tau_{21}\rangle$ is reached around $z \sim 10$, when the universe
is a few percent ionized. To change the average ionization fraction corresponding to the maximum it would be necessary to
modify the ratio between the X-ray photons production rate and the Lyman-band photons production rate. Changing $f_X$ or the IMF would
achieve this. For example a lower $f_X$ would shift the maximum to a lower redshift and a larger value. We find that model 2 shows a behaviour similar in shape and amplitude to the semi-analytical work by \citet{Furlanetto06c} that  includes a simple modelling of the Wouthuysen-Field coupling, albeit shifted to lower $z$ in our case. This is due to a later star formation history imposed by the limited resolution of the simulation. Model 1 is similar to the results of \citet{Mack12} who do not model the Wouthuysen-Field coupling, but shows a discrepancy (difference in amplitude at high $z$) with \citet{Ciardi13} who include the effect of Ly-$\alpha$ photons for heating but not for Wouthuysen-Field coupling.

Although the volume-averaged value of $\tau_{21}$ is the first and simplest indicator of the absorption level in the spectrum of radio-loud background sources, it tell us very little about the observability of the phenomenon. When analysing the data, the average absorption will be fitted out along with the continuum. Only under the assumption
that, along the line of sight, the absorption shows fluctuations of amplitude similar to its average value (which is true for example if an ionized bubble intersects the line of sight) does it begin to be relevant to assess the observability. A more telling quantity is the largest expected absorption value in the spectrum for a given frequency range. Considering that the spectral resolution for this type of observations on such instruments as LOFAR or SKA
will be a few kHz and that we would like to trace the ionization history with a resolution better than
$\Delta z =1$, that is a frequency range of a few MHz, it is reasonable to estimate the average value of the  
one-thousandth cells in the simulation box with the largest $\tau_{21}$ values. This is the quantity that
is plotted, for the different levels of modelling, in the lower panel of Fig. \ref{average_tau}. It shows, of course,
an overall upward shift of the curves, but more importantly it shows that for this population of cells producing large
absorption features, the effect of including peculiar velocities increase the optical depth by a factor of $\sim 2$.

\subsection{Impact of the physical modelling}
Before presenting more detailed statistics of the distribution of $\tau_{21}$ values, we would like to highlight the impact
of three non-trivial physical processes on the predicted values.

\begin{figure*}
\begin{tabular}{cc}
\includegraphics[width=8.5cm]{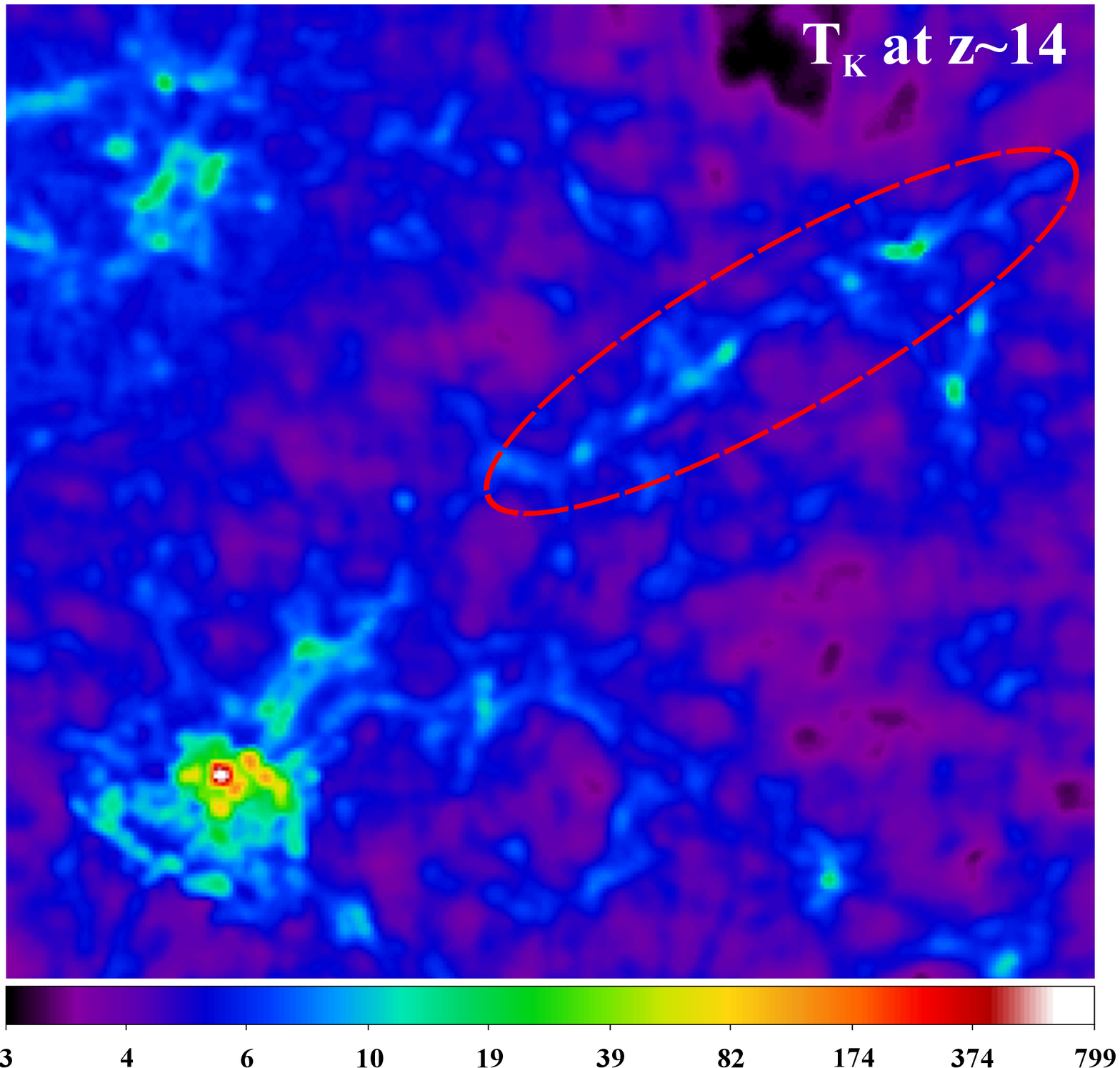} &
\includegraphics[width=8.5cm]{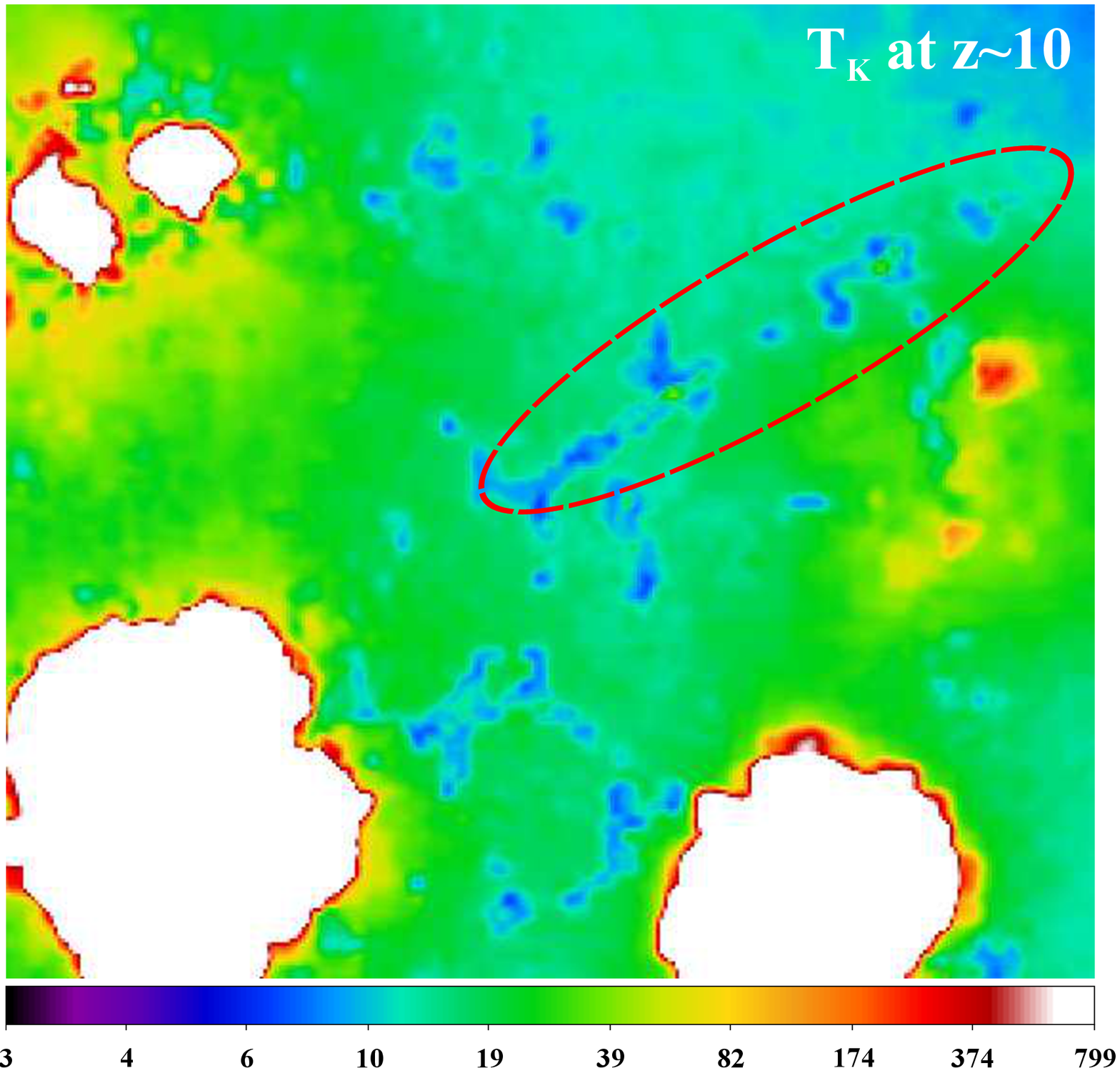} \\
\includegraphics[width=8.5cm]{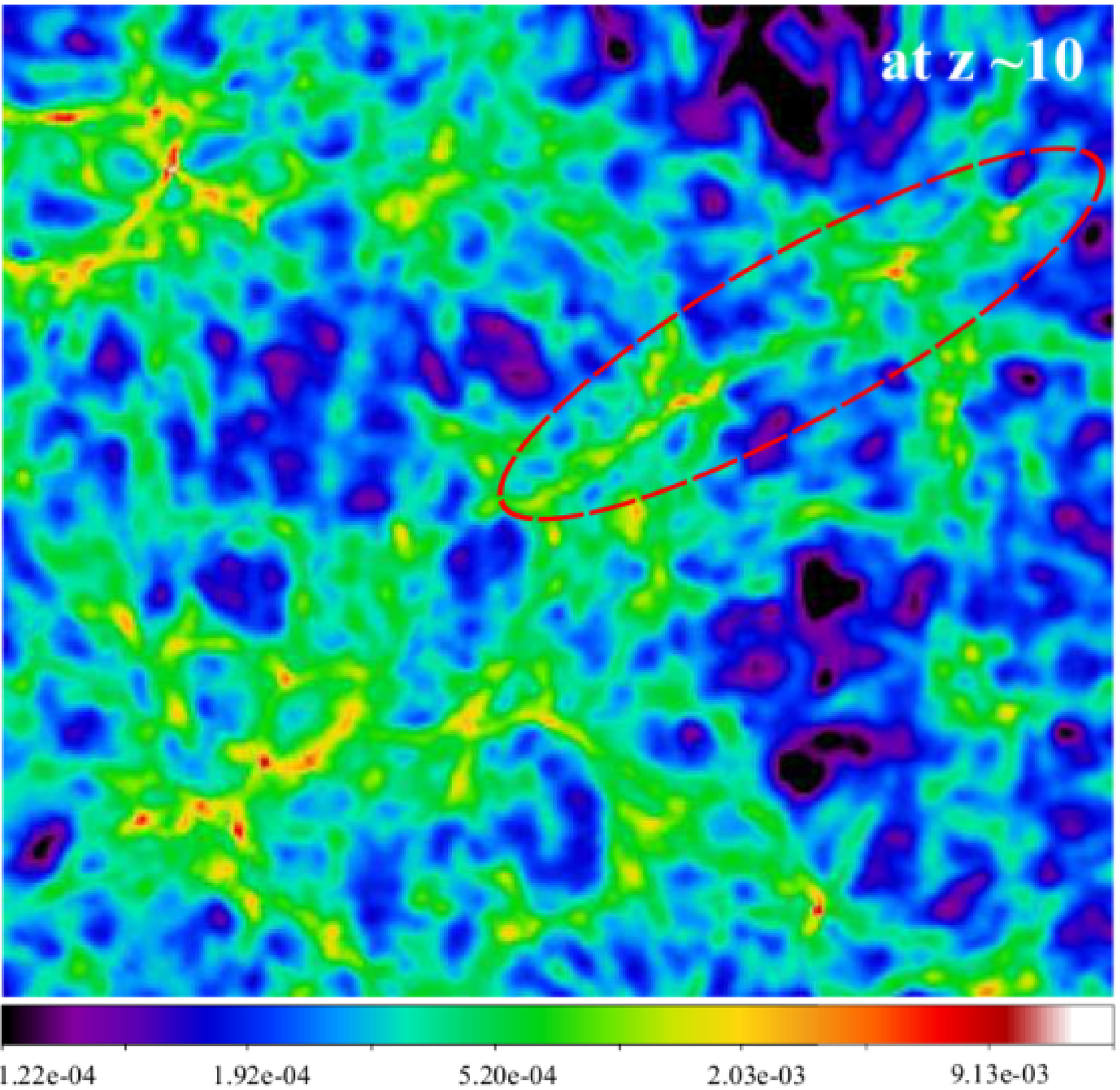} &
\includegraphics[width=8.5cm]{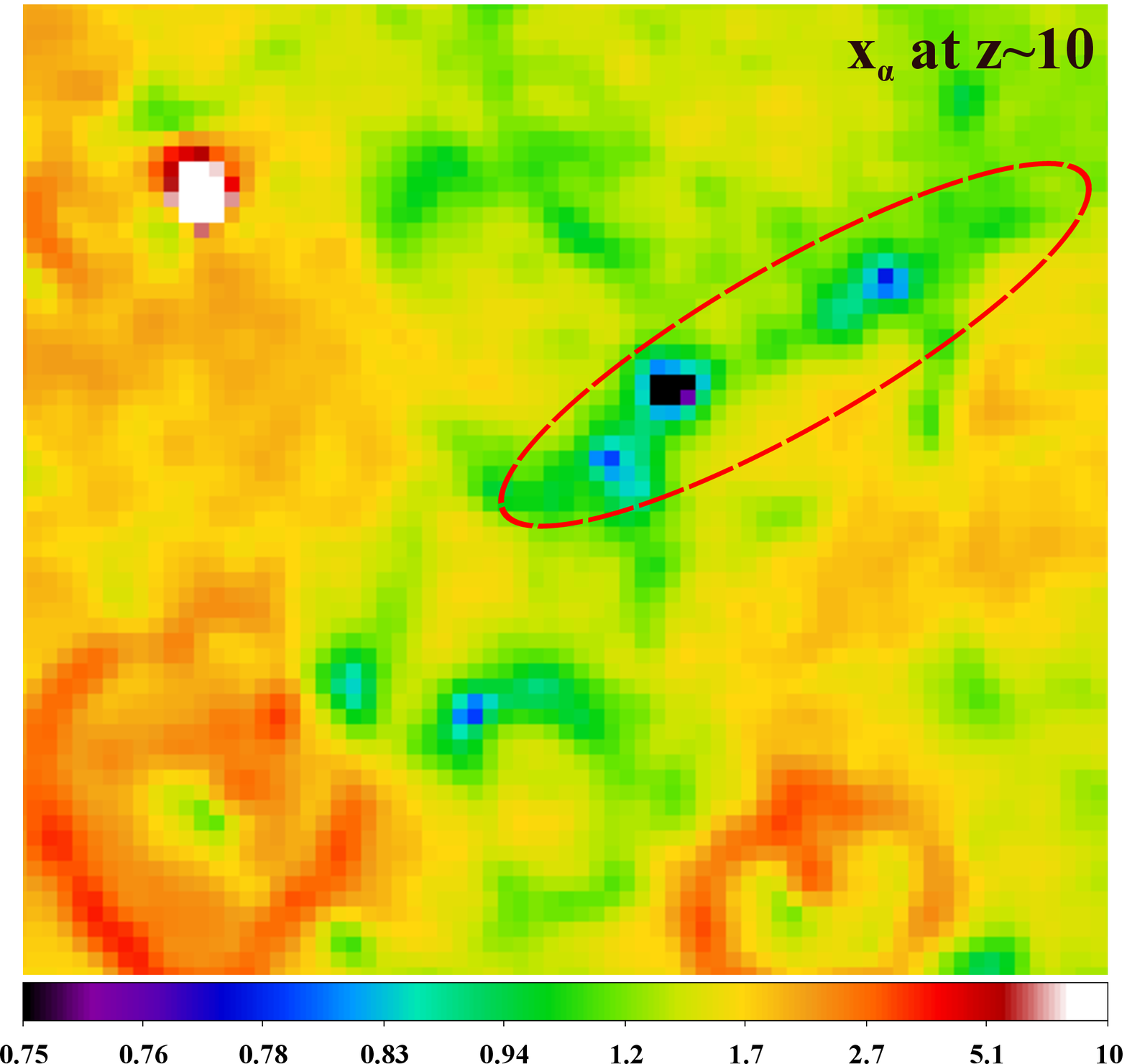} 
\end{tabular}
\caption{Maps of a region $\sim 10$ cMpc on a side. The thickness of the slice is $35$ ckpc for the first three and $140$ ckpc for the bottom right map. The plotted quantities are: kinetic temperature at $z \sim 14$ (top left), kinetic temperature at $z \sim 10$ (top right), number density in cm$^{-3}$  at $z \sim 10$ (bottom left) and $x_\alpha$ coefficient at $z\sim 10$ (bottom right).}
\label{X-ray}
\end{figure*}

\subsubsection{Partial self-shielding from soft X-rays}

It is generally considered that X-ray travel long distances in the neutral IGM before interacting with hydrogen atoms. This however
is every dependent on the energy of the photons. Indeed the comoving mean free path is $l=2. (1+ \delta)^{-1} \left({E \over E_0}\right)^3\left({10 \over 1+z}\right)^{2}$ ckpc, where $\delta$ is the overdensity, $E$ the energy of the photon, and $E_0$ the energy at the ionization threshold. If $E= 2$ kev the mean free path approaches the Hubble radius, but if $E=100$ eV, the mean free path at overdensities of a few is of the order $100$ ckpc.
In the case of a soft X-ray spectrum a large fraction of the energy is emitted below $300$ eV ($90 \%$ for a 1.6 spectral index). Consequently we can expect that a structure with an overdensity of a few extending over $\sim 1$ cMpc at $z \sim 10$ will have a non-negligible optical thickness and will be partially self-shielded from a soft X-ray flux. As a result X-rays will heat it less than the surrounding diffuse gas. 

This effect is visible in Fig. \ref{X-ray}. In the upper left panel, where X-ray heating has not occurred yet, the temperature of the high-density region within the red ellipse is higher than that of the lower density surrounding medium because of adiabatic contraction. In the upper right panel, after
X-ray heating kicked in, the temperature contrast is inverted. The low-density regions have been heated more efficiently, their temperature rising above that of the high-density regions. Structures
exhibiting X-ray self-shielding are ubiquitous in the simulation box.
It is however necessary to resolve the radiative transfer down to scales $\sim 100$ ckpc to capture this effect with reasonable accuracy. The resolution in our simulation is marginally adequate.

In terms of $\tau_{21}$, these self-shielded regions that have optical depth higher than average due to their density will receive
a further boost from not being efficiently heated by X-rays.

\begin{figure}
\includegraphics[width=8.3cm]{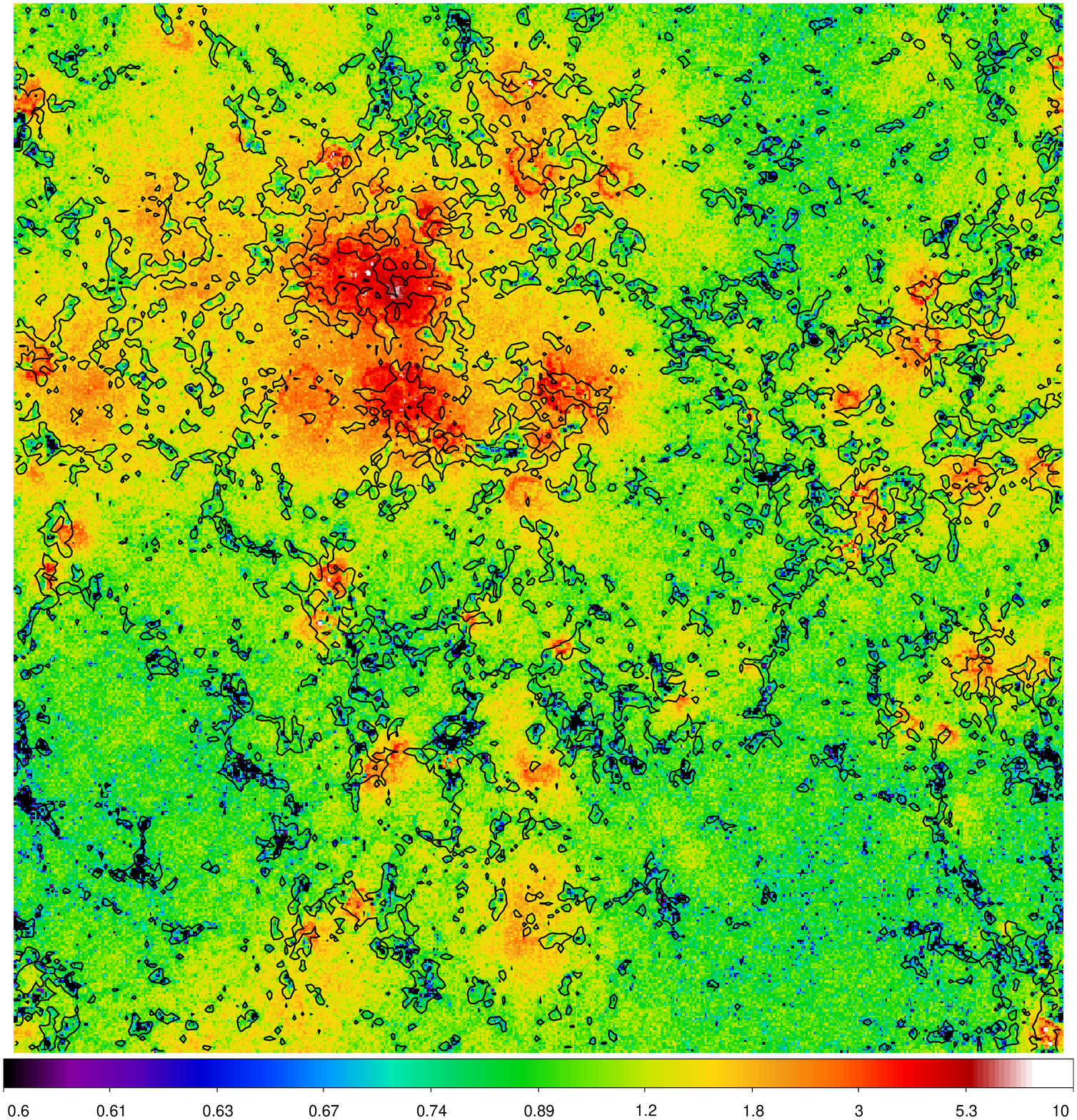} 
\caption{Map of $x_\alpha$ in a $140$ ckpc thick slice of the whole box corresponding to a single cell of the Lyman-$\alpha$ transfer simulation, at the redshift of strongest absorption ($z=10.1$). No smoothing a been applied to the data. The overlaid contour is derived from the density field and corresponds to $1.38$ times the average density of the universe.}
\label{x_a}
\end{figure}

\subsubsection{Partial self-shielding from Ly-$\alpha$ pumping}
The radiative transfer of photons in the Lyman lines is more complex than that of ionizing photons for two reasons.
First, photons are most often scattered in the Lyman lines, not absorbed. Second, the Lyman-$\alpha$ line is strong enough in typical IGM conditions during the EoR that scattering in the wings of the line does occur. In practice, a photon redshifting toward the Lyman-$\alpha$ frequency will be scattered for the first time in the wing of the line $\sim 10$ cMpc from the location where it would reach the local rest frame Lyman-$\alpha$ frequency \citep{Semelin07}. The first order effect
of wing scatterings is that for a flat-spectrum source the radial Lyman-$\alpha$ flux profile does not follow a $r^{-2}$ law as would be the case for a Dirac-like line profile. Even in a homogeneous IGM, there is a steepening of the profile close to the source ($< 10 $ cMpc for the IGM at $z\sim 10$) to $\sim r^-{7 \over 3}$ \citep{Chuzhoy07,Semelin07}. A more subtle effect, shown in \citet{Semelin07} in an idealized situation, is that the local Lyman-$\alpha$ flux may be depleted in moderately overdense regions. Photons that should redshift to the local frame Ly-$\alpha$ frequency at the centre of the overdense structure are instead scattered in the wing of the line with increased probability in the outer parts of the overdense structure and may bounce off the structure altogether. 

This effect is visible in the lower right panel of Fig. \ref{X-ray} where $x_a$, which is proportional to the local Lyman-$\alpha$ flux, is depleted in overdense regions. Such an effect was not clearly detected in previous works \citep{Baek09,Vonlanthen11}  because it requires a spatial resolution of a few 100 ckpc at least and a low Monte Carlo noise level. It would be near impossible 
to see it in a $200$ cMpc simulation box with the current computational power and even in our box it required a large number of photons. The depletion in $x_a$ reaches a factor $2-3$ at overdensities of $\sim 10$ where collisional coupling is not quite sufficient to compensate. Fig. \ref{x_a} shows a map of $x_\alpha$ values for a thin slice of the whole simulation box. 
Numerous dark regions associated with low $x_\alpha$ values are enclosed in a contour showing regions with a density larger than 1.38 times the average density of the universe. This shows that the partial self-shielding from Ly-$\alpha$ is ubiquitous in the neutral IGM.

Except in scenarios with very strong X-ray heating, $T_K$ will be lower than $T_{\mathrm{CMB}}$ at redshifts when the
Wouthuysen-Field coupling through Ly-$\alpha$  is not saturated and thus the self-shielding effect most significant. 
Consequently the hydrogen spin temperature $T_s$ will be higher (closer to $T_{\mathrm{CMB}}$) in overdense regions and
$\tau_{21}$ lower than in the absence of self-shielding. This effect, that affects the same regions as the previous one,
goes in the opposite direction but is relevant during a shorter period.

\subsubsection{Gravitational (shock) heating}
It is usually considered that coupled radiative hydrodynamics is needed in simulating the universe during the EoR only
if structures with masses below $10^8$ M$_\odot$ are resolved, as their self-gravity is not sufficient to resist photo-heating. Focussing on moderately overdense structures in the neutral IGM however, the two relevant heating processes are
heating by X-rays and heating through hydrodynamic dissipation during gravitational collapse. If these two are not computed simultaneously, it is not possible to compute $T_K$ accurately. Dissipation should not be very strong in moderately overdense structures, in the early stages of gravitational collapse. One the other hand, even heating of
a few kelvins is significant for computing $\tau_{21}$.  It is thus reasonable to check the impact of gravitational heating on
$\tau_{21}$ and whether running radiative hydrodynamics simulations is needed.

\begin{figure}
\includegraphics[width=8.3cm]{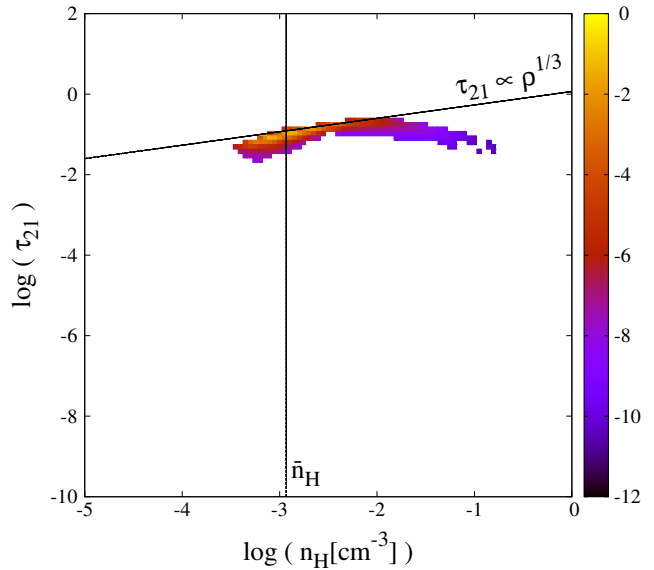} 
\caption{Distribution of $\tau_{21}$ values as a function of density at $z\sim 15$ before any ionizing or X-ray photons have been emitted. $T_s=T_K$ is assumed. The colour scale is the normalized logarithmic abundance. The vertical line shows the average baryon density and the slanted line show the maximal $\tau_{21}$ if $T_K$ has evolved adiabatically (see main text). Departure from this line shows that heating occurred (through shocks).}
\label{shocks}
\end{figure}

If we compute $\tau_{21}$ assuming  $x_i=0$, $T_s=T_K$ (which is inconsistent) and neglecting the velocity effect, formula (1) simplifies to $\tau_{21}=0.0092 (1+z)^{3 \over 2} {1+\delta \over T_K}$. If we further assume that the gas followed an adiabatic evolution from $z \simeq 150$ when it thermally decoupled from the CMB, without significant heating or cooling, and that density fluctuations were negligible at $z=150$ compared to $z=6-15$, we get the simple relation $\tau_{21} \simeq 0.16 \left({10 \over 1+z}\right)^{1 \over 2}(1 + \delta)^{1 \over 3} $. In some sense, this the
largest possible absorption value (not including velocity effect). In Fig. \ref{shocks} we compare the value found in
the simulation before the formation of the first source (computed using $T_s=T_K$ to conform to the above assumptions) with this maximal-absorption relation. We find that at overdensities a few tens, the simulated values are lower than dictated by the maximal absorption relation. We deduce that the corresponding temperatures are higher than those obtained through pure adiabatic evolution. No X-ray having been produced yet, the only possible cause is gravitational heating that is thus
revealed to have a significant impact on the $\tau_{21}$ in overdense regions. If radiative transfer
is run as a post-treatment it will be difficult to combine accurately X-ray heating, adiabatic heating and shock heating. 

It is reasonable to mention that correctly modelling gravitational heating in any kind of cosmological simulation is a challenge. In reality
this heating probably proceeds through the formation of shocks on scale much smaller than the resolution of the simulation, where the kinetic energy of the gas is radiated away. While there is little doubt that kinetic energy needs
to be dissipated (otherwise shell crossing would occur unimpeded and galaxies would not condense), it is less clear whether radiations will redistribute the energy uniformly in the form of atomic thermal motion in the surrounding medium and on what scale. The viscosity of the SPH simulation ignores these complex processes and simply dissipates the energy at the simulation resolution. Consequently, while we believe that gravitational heating cannot be simply ignored in detailed
modelling of the $21$-cm forest, we recognize that its quantitative effect remains a source of uncertainty.
 
\subsection{Statistical properties of the optical depth values}

It is now time to characterize the distribution of $\tau_{21}$ in more details than the simple average value. We will show that the distribution is clearly non-Gaussian. Thus the average and variance do not encode the full information
concerning the distribution. We will study the full Probability Distribution Function (PDF).

\subsubsection{ The PDF of $\tau_{21}$}
First, let us remember that the thermal width of the $21$-cm line can be converted through the Hubble flow to scales of the order $10$ ckpc in the typical conditions of the neutral IGM during the EoR. Travelling this distance, a photon will redshift from one side of the line to the other. This scale will act as a smoothing length on the $21$-cm forest which is a continuous superposition of such lines.
An observation of the $21$-cm forest with spectral resolution $\Delta \nu$ will resolve scales of $\sim 17. \left({\Delta \nu \over 1 \mathrm{kHz}} \right) \left({1+z \over 10}\right)^{1 \over 2}$ ckpc. Thus the SKA, with an expected spectral resolution of 1-5 kHz for $21$-cm forest observation, will recover most of the information in the signal (provided the background source is bright enough). The
cell size in our simulated $\tau_{21}$ cube is $\sim 35$ ckpc, in the same range as the SKA spectral resolution and the typical thermal width of the line. This does not mean that a higher resolution simulation would not change the
simulated PDF but it would also require a more complex computation of $\tau_{21}$ to account for variation of physical
quantities within the thermal  line width. 

\begin{figure}
\includegraphics[width=8.3cm]{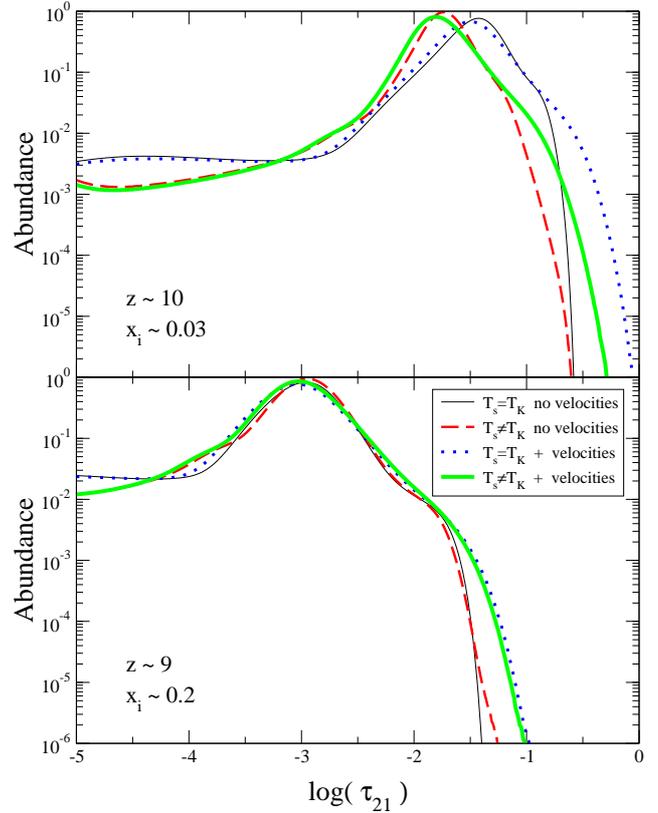}
\caption{Probability distribution function of $\tau_{21}$ computed from the simulation date at the redshift of strongest average absorption (upper panel) and when the average absorption has decreased by a factor of 10 (lower panel). The different curves correspond to the levels of modelling listed in table 1. }
\label{PDF}
\end{figure}

The PDF of $\tau_{21}$ is presented in Fig. \ref{PDF} at $z\sim 10$ when the
average value of $\tau_{21}$ is largest and at $z \sim 9$ when the average of $\tau_{21}$ has decreased by a factor of $\sim 10$, mainly due to X-ray heating. First we can check that the PDF is indeed non-Gaussian. Next, and this is the
crucial information in this figure, we see that, at $z \sim 10$, the full computation of $T_s$ affects a large range of
$\tau_{21}$ value, while including peculiar velocity effects alters mainly the large $\tau_{21}$ wing of the distribution
which is consistent with the lower panel of Fig. \ref{average_tau}. Peculiar velocities enhance the PDF at $\tau_{21} \gtrsim 3\langle\tau_{21}\rangle$. Later on, as the Wouthuysen-Field effect  saturates, the $T_s=T_K$ approximation does not change the PDF (but the prospects of detection are much reduced) and including peculiar velocities in the computation only
changes the distribution for the rare cells at  $\tau_{21} \gtrsim 20 \,\langle\tau_{21}\rangle$.

\begin{figure*}
\includegraphics[width=17cm]{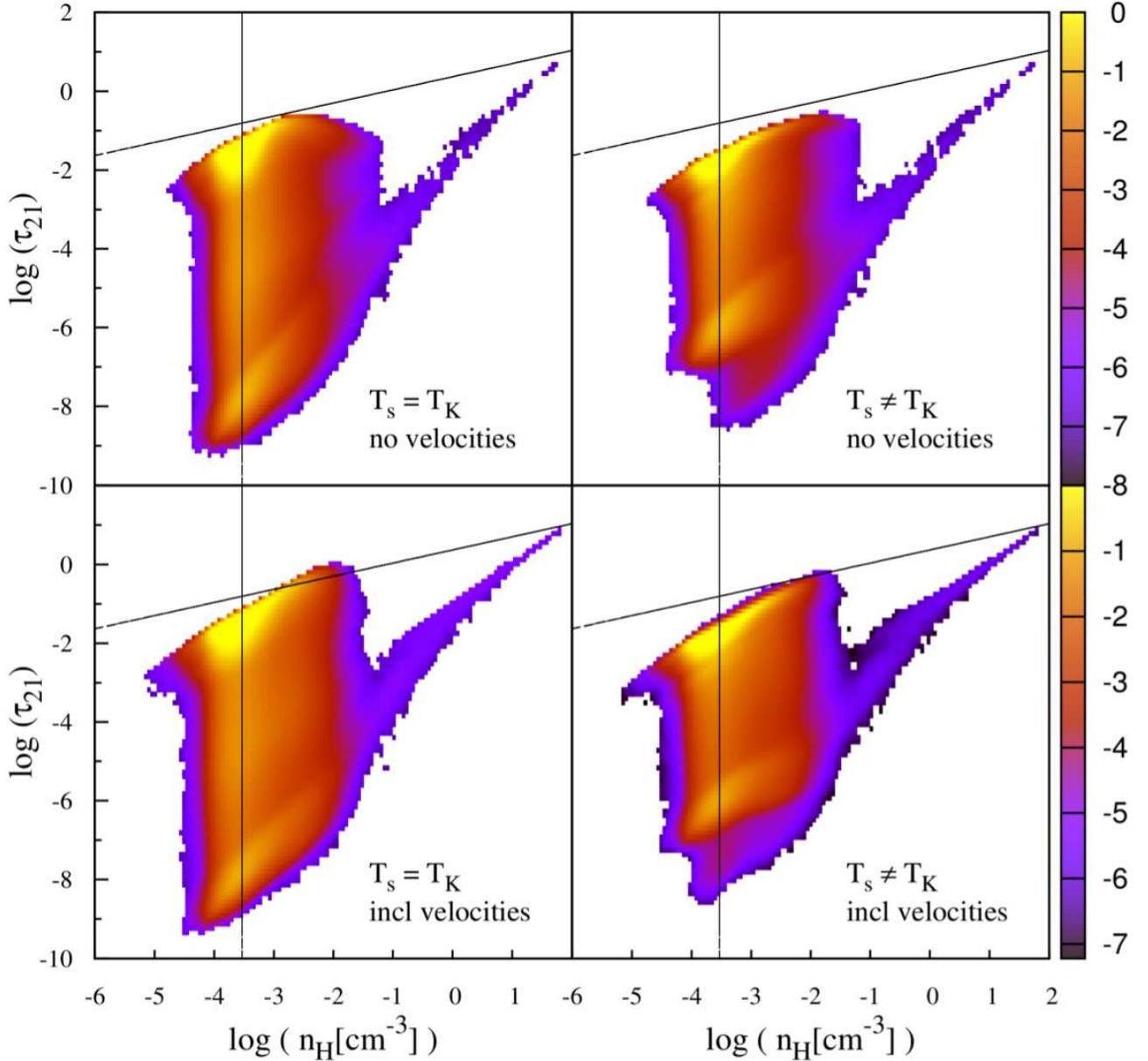}
\caption{Distribution of $\tau_{21}$ values as a function of density at the redshift of strongest average absorption. The lines have the same meaning than in Fig. \ref{shocks}. The different panels corresponds to
the different levels of modelling listed in table 1; the associated assumptions are explicit in the panels.
The colour codes the normalized logarithmic abundance of cells in the bin.}
\label{2Dhist}
\end{figure*}

\subsubsection{Distribution of $\tau_{21}$ as a function of the local density}
$\tau_{21}$ is a function of several local quantities, namely the hydrogen atom number density, the velocity, the
local Lyman-$\alpha$ flux and the kinetic temperature. The Lyman-$\alpha$ flux and kinetic temperature have large scale
fluctuations, seeded by the distribution of sources, that are uncorrelated with the local density fluctuations in the IGM. Their small scale fluctuations through self-shielding and adiabatic evolution, on the other hand, are very much correlated with the local density. In the same way, peculiar velocity gradient are strongly correlated with the local density fluctuations. It appears then that the local baryon density is a crucial quantity in determining $\tau_{21}$. 

In Fig. \ref{2Dhist} we show the distribution of $\tau_{21}$ has a function of the total hydrogen number density at
$z\sim 10$ for four different levels of approximation. Let us first mention that the high density ($n_\mathrm{H} \gtrsim 3. \, 10^{-2}$ cm$^{-3}$) extension of the distribution is associated with gas particles at densities larger than the star formation threshold located in primordial galaxies, not in the IGM. While subject to a strong ionizing background, they
are dense enough to retain some neutral hydrogen due to recombination. We do not consider that the resolution of our simulation is good enough to model the interstellar medium robustly and we do not further interpret this feature of the distribution. 

The interesting part of the distribution is where $\tau_{21} > 10^{-2}$, in moderately overdense or underdense regions ($10^{-4} \lesssim n_\mathrm{H} \lesssim 10^{-2})$. In the upper left panel, we see how $\tau_{21}$
is prevented from reaching the maximum value derived from a pure adiabatic evolution for the gas ($\tau_{21} \propto \rho^{1 \over 3}$, materialized by the slanted line) by X-ray heating at densities around and below the average density
and by shock heating at overdensities larger than $\sim 10$. We are guided in this interpretation by Fig. \ref{shocks} where no X-ray heating has occurred. The upper right panel shows how taking into account the unsaturated Wouthuysen-Field coupling results in an overall decrease in $\tau_{21}$, except at overdensities of a few tens where $\tau_{21}$ actually increases. Indeed, at such densities Lyman-$\alpha$ self-shielding bring $T_s$ closer to $T_\mathrm{CMB}$ which is actually lower than the adiabatically increased $T_K$. In the lower left panel we see that including peculiar velocities
in the computation of $\tau_{21}$ boosts the values by a factor of a few in overdense collapsing structures. This effect
has been studied before, mainly in the context of the $21$-cm 3D signal \citep[e. g.][]{Mellema06,Mao12}. It is especially
relevant for the $21$-cm forest, boosting the signal in the regions that will be easiest to detect and thus making them detectable with dimer background sources. Finally the lower right panel presents the most accurate modelling, including
both the effect of peculiar velocities and the correct computation of $T_s$. It is worth noticing that, with our best
modelling, we find maximum $\tau_{21}$ values a factor of a few larger than with the more approximate modelling in the upper left panel, that has been often used in previous works. This, of course, improves the prospect for observational detection.

\begin{figure}
\includegraphics[width=8cm]{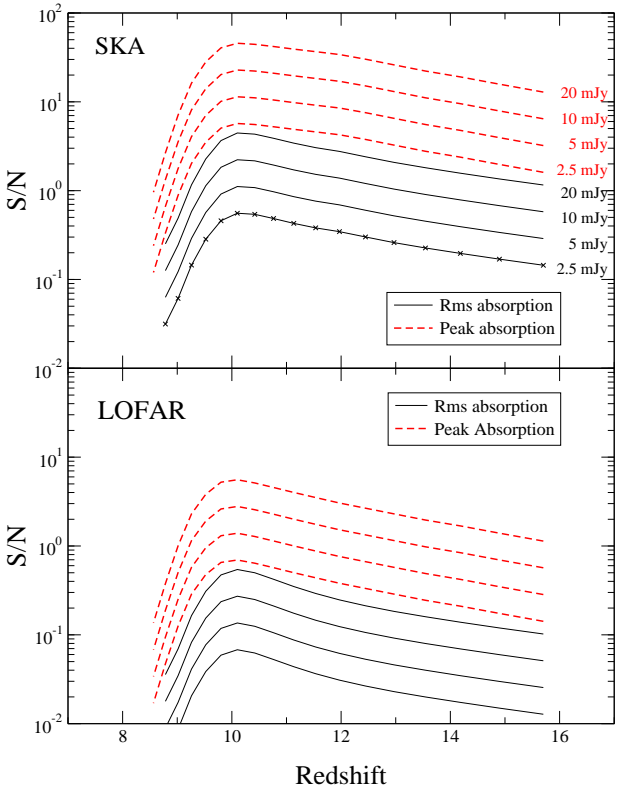}
\caption{Expected signal-to-noise ratio for the measurement of the $21$-cm forest against background radio sources of different luminosities as a function of redshift. The crosses indicate the redshifts of the snapshots where the signal-to-noise was computed; the curves are interpolations between these points. The radio source luminosity is independent of redshift and consequently of observed frequency. Typical rms absorption is plotted in back and the peak absorption in dashed red (see main text for definition). Top panel is for the SKA, bottom panel for LOFAR. Luminosities are the same in both panels. }
\label{SN-luminosity}
\end{figure}

\subsection{Prospects for detection}

Estimating the observability of the $21$-cm forest during the EoR is not the main focus of this work. However, we showed
that several aspects of a careful modelling have an impact on the predicted value of the optical depth especially in
high value regions. The combined result seems to be an increase of a factor of a few for the largest values of the optical depth. We will now present simple estimations of the expected signal-to-noise ratio for observations with LOFAR and SKA of our modelled signal.

\subsubsection{Strongest and typical absorption feature detection}
We will base our analysis on the whole population of $\tau_{21}$ values in each snapshot of the simulation ($2048^3$ values) that give us much better statistics than using a single line of sight. A single cell corresponds to an observed
signal $S_0 e^{-\tau_{21}}$, where $S_0$ is the unabsorbed spectral flux received from the background source. However we deal with a continuous (but fluctuating) absorption along the line of sight, so after the subtraction of the average absorbed continuum from the source, the relevant signal is $S_0 (e^{-\tau_{21}}-\langle e^{-\tau_{21}} \rangle)$. Thus
the rms value of this quantities is a significant statistical quantity to characterize the signal. We will refer to it simply
as the \textit{rms absorption}. It allows us to quantify the possibility to extract most of the information along the line of sight in an observed spectrum. Another relevant diagnostics is the possibility to detect the largest absorption feature in a given frequency or redshift range. Here we will use again, as in section 3.1, a frequency range of $10$ Mhz and spectral resolution of 10 kHz, and thus we define the \textit{peak absorption} as the maximum of the 1/1000th smallest $e^{-\tau_{21}}$ values, multiplied by $S_0$. We will compare the rms and peak absorption to the noise level.

We model the observational aspects with a very simple approach. First we will consider that the background source
is dominant in the beam of the radio telescope, neglecting the effect of diffuse foregrounds and other point sources in the main lobe or in side lobes. Of course, this works best with a radio interferometer with long base lines and a tight beam. We will ignore issues related to the ionosphere, radio-interference and calibration. Consequently we simply
model the level of noise with the radiometer equation:

\begin{equation}
\sigma_n= {1 \over \eta} {2 k_B T_{\mathrm{sys}} \over A_{\mathrm{eff}} \sqrt{\Delta \nu t_{\mathrm{int}}}}
\end{equation}
where $T_{\mathrm{sys}}$ is the system noise including the electronics and the sky temperature, $T_{\mathrm{sys}}= 100+ 300\left({\nu[\mathrm{MHz}] \over 150 }\right)^{-2.55}$ K \citep{Mellema13}, $k_B$ is the Boltzmann constant, $A_{\mathrm{eff}}$ is the total effective collecting area of the radio telescope, $\Delta \nu$ is the frequency resolution, $t_{\mathrm{int}}$ is the integration time and $\eta$ an overall efficiency coefficient for the instrument that we set equal to $0.5$. Following \citet{Ciardi13} we set $A_{\mathrm{eff}}=48 \times 24\times 16 \times \mathrm{min}({\lambda^2 \over 3}, 1.56)$ for LOFAR and for the SKA we consider the current rebaselining guideline for $130\,000$ dipoles and an optimal frequency of $108$MHz, that is  $A_{\mathrm{eff}}= 130\,000 \times \mathrm{min}({\lambda^2 \over 3}, 2.56)$. We consider an integration time of $t_{\mathrm{int}}=1000$h and a fiducial spectral resolution $\Delta \nu = 10$ kHz.

In Fig. \ref{SN-luminosity}, the resulting signal-to-noise ratio of both rms and peak absorption is plotted for the SKA (top panel) and for LOFAR (bottom panel) for sources with varying luminosity. Each points of each curve is computed from statistics over all the pixel of the snapshot corresponding to the redshift value. Obviously, a single line of sight observation will not benefit from such high statistics and may show some variance compared to this plot. The evolution of the signal and of the noise with redshift is fully taken into account. 

From this plot we find that the SKA can measure most of the absorption information only for $\sim 20$ mJy background sources and for a slice of the IGM 
during the early EoR (no more than a few percent of ionization). If this period in the history of the EoR is shifted
to higher redshift, the rms absorption would be more difficult to measure due to the increased sky temperature. In contrast, the peak absorption could be detected for sources as dim as $2.5$ mJy located at high enough redshift. With LOFAR, using our simple modelling, only the peak
absorption should be detectable for $\sim 20$ mJy sources. If X-ray heating of the IGM is less efficient than in our
simulation, the observational prospects improve somewhat. Finally let us emphasize again that minihaloes are not resolved in our simulation. As they may be responsible for strong absorption features, including them could improve the signal-to-noise for both the peak and rms absorption.

\begin{figure}
\includegraphics[width=8cm]{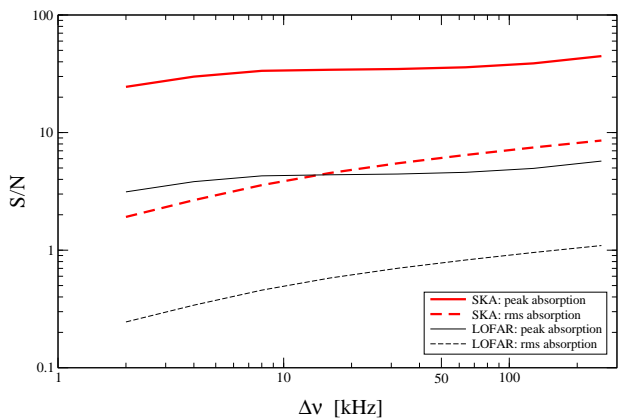}
\caption{Signal-to-noise ratios as a function of channel width for the observation of the $21$-cm forest a $z\sim 10$ against a 20 mJy background radio source. The ratios for the rms absorption are in dashed lines while the ratios for peak absorption are in full lines. Ratios for the SKA are in thick red, ratios for LOFAR are in thin black.}
\label{SN-bandwidth}
\end{figure}

\subsubsection{ Influence of the spectral resolution}
A relevant question is: what spectral resolution should be used to optimize the detection of the signal? If a coarser resolution is used the noise level decreases, but the signal from an absorption feature narrower than the spectral resolution will be diluted in the channel. In the limit of a dominant and very narrow absorption feature, the signal-to-noise ratio drops as $\Delta \nu^{-{1 \over 2}}$. On the opposite, for a signal with spectral fluctuations on scales much larger than the resolution, the signal-to-noise increases as $\Delta \nu^{1 \over 2}$. Using a spherically symmetric analytical model, it has been
shown that minihaloes are expected to produce absorption features on scales of a few kHz \citep{Furlanetto02,Xu11,Meiksin11}. Using a resolution of a few tens of kHz would then smooth out the signal from minihaloes. Milder overdensities such a filaments also produce absorption features. They are less symmetric but easier to study with numerical simulations because they require lower mass resolution. \citet{Ciardi13}, for example, find that smoothing the signal on $20$ kHz scale, a procedure similar to using a $20$ kHz spectral resolution, improves the prospects for detection of strong absorption features.

We rebinned our $\tau_{21}$ data cubes in blocks of $1\times1\times2^n$ cells ($n=1,2,...,8$) corresponding to $8$ logarithmically spaced frequency resolution values between $\Delta \nu=2$ to $256$ kHz and re-evaluated the typical peak and rms absorption in a $10$ MHz frequency range for each spectral resolution. Note that as frequency resolution worsen, the number of resolution elements in the $10$ MHz frequency range decreases, and thus the typical peak absorption is evaluated using a fraction of the blocks of cells growing as ${\Delta \nu \over 10 \mathrm{ MHz}}$. In Fig. \ref{SN-bandwidth} we
plot the signal-to-noise ratios as a function of $\Delta \nu$ for peak and rms absorption of a $20$ mJy background source observed with LOFAR and the SKA at
$z\sim 10$. We can see than the detectability of the peak absorption features does not drop much toward small $\Delta \nu$ indicating that these features
occurs on a few kHz scale (which is confirmed by looking at the synthetic spectra themselves), corresponding  a few tens of comoving kpc in terms of 
spatial scales. Let us mention that the average inter-particle distance in these mildly overdense structures is still more than 10 times larger than
the gravitational softening $\epsilon = 1.5 $ ckpc: these are well resolved scales but a finer mass resolution might overlay even smaller scale structures 
on top of them. With increasing $\Delta \nu$, the signal to noise does not vary much over a large range of frequency resolution, invalidating
the hypothesis of a single, isolated and dominant absorption feature that would follow a $S/N \propto  \Delta \nu^{-{1 \over 2}}$ law at large $\Delta \nu$.

The rms absorption behaves somewhat differently,
showing a steady increase in S/N toward larger $\Delta \nu$, indicating 
a substantial contribution of large scale fluctuations in $\tau_{21}$ up to
scales of several cMpc ($> 100$ kHz). These behaviours may be used as guidelines in
the analysis of observational data that should be recorded at the finest possible instrumental resolution anyway.
\section{Conclusions}

We have performed radiative hydrodynamics simulations in a cosmological volume to model the $21$-cm forest. X-ray heating and Wouthuysen-Field coupling are evaluated from 3D radiative transfer. Our comparatively high resolution for radiative transfer (of the order $10$ ckpc) and the choice of a large
number of photon packets to reduce the Monte Carlo noise have allowed us to reveal that moderately 
overdense structures are partially self-shielded from both the (soft) X-ray background and the
Lyman-$\alpha$ flux. Both effects modify the $21$-cm optical thickness by a factor of a few (in opposite directions). The net effect depends on redshift (Lyman-$\alpha$ saturates while X-ray self-shielding
keeps limiting the temperature increase) and on the relative production rate of Lyman-band photons 
and X-ray photons as a function of $z$. We have also shown that shock heating cannot be ignored in these moderately overdense collapsing structures and modifies $\tau_{21}$ by a factor of a few. 

Looking at the distribution of $\tau_{21}$ values as a function of the local density, we examine the impact of including a complete computation of $T_s$ and the effect of peculiar velocities (along with the previously mentioned self-shielding and heating effects). The most relevant net trend is to boost
the $\tau_{21}$ value by a factor of the few in structures with overdensity of the few tens. Finally
we evaluate that the latest re-baselined SKA would be able to extract most of the information in the 
spectrum of a $20$ mJy background source at $z\sim 10$ with a $1000$h integration and would be able to detect a few absorption features if the luminosity of the source is $2.5$ mJy. A larger channel width is preferable when going after the rms absorption level (possible in the case of a luminous source), while the S/N is quite insensitive to the channel width (at least up to $256$ kHz) when trying to detect absorption peaks only (the only option for a less luminous source).

The halved collecting area of SKA-Low following the recent re-baselining process (130 000 dipoles), combined with the small number of currently detected bright radio-sources at $z>6$ strongly argue in
favour of a multibeaming capacity for SKA-Low. There is a high probability that suitable sources for
$21$-cm forest measurement will lie outside of the chosen instantaneous $21$-cm tomographic survey field. As deep integration is more or less required, tomographic survey and $21$ forest measurement would ideally be performed simultaneously.

\section*{Acknowledgments}I would like to thank B. Ciardi, F. Combes, and C. Tasse for useful discussions. I would also like to thank F. Marcadon during whose internship we made our first forays into the $21$-cm forest simulation. 
This work was made in the framework of the French ANR funded project ORAGE (ANR-14-CE33-0016). We also acknowledge the support of the ILP LABEX (under the reference ANR-10-LABX-63) within the Investissements
d'Avenir programme under reference ANR-11-IDEX-0004-02. The simulations were performed on the GENCI
national computing center at CCRT and CINES (DARI grants number 2014046667 and 2015047376).
 \bibliographystyle{mnras}
\bibliography{myref}

\begin{thebibliography}{}
\makeatletter
\relax
\def\mn@urlcharsother{\let\do\@makeother \do\$\do\&\do\#\do\^\do\_\do\%\do\~}
\def\mn@doi{\begingroup\mn@urlcharsother \@ifnextchar [ {\mn@doi@}
  {\mn@doi@[]}}
\def\mn@doi@[#1]#2{\def\@tempa{#1}\ifx\@tempa\@empty \href
  {http://dx.doi.org/#2} {doi:#2}\else \href {http://dx.doi.org/#2} {#1}\fi
  \endgroup}
\def\mn@eprint#1#2{\mn@eprint@#1:#2::\@nil}
\def\mn@eprint@arXiv#1{\href {http://arxiv.org/abs/#1} {{\tt arXiv:#1}}}
\def\mn@eprint@dblp#1{\href {http://dblp.uni-trier.de/rec/bibtex/#1.xml}
  {dblp:#1}}
\def\mn@eprint@#1:#2:#3:#4\@nil{\def\@tempa {#1}\def\@tempb {#2}\def\@tempc
  {#3}\ifx \@tempc \@empty \let \@tempc \@tempb \let \@tempb \@tempa \fi \ifx
  \@tempb \@empty \def\@tempb {arXiv}\fi \@ifundefined
  {mn@eprint@\@tempb}{\@tempb:\@tempc}{\expandafter \expandafter \csname
  mn@eprint@\@tempb\endcsname \expandafter{\@tempc}}}

\bibitem[\protect\citeauthoryear{{Afonso}, {Casanellas}, {Prandoni}, {Jarvis},
  {Lorenzoni}, {Magliocchetti}  \& {Seymour}}{{Afonso} et~al.}{2015}]{Afonso15}
{Afonso} J.,  {Casanellas} J.,  {Prandoni} I.,  {Jarvis} M.,  {Lorenzoni} S.,
  {Magliocchetti} M.,   {Seymour} N.,  2015, in Advancing Astrophysics with the
  Square Kilometre Array (AASKA14). p.~71 (\mn@eprint {arXiv} {1412.6040})

\bibitem[\protect\citeauthoryear{{Ba{\~n}ados} et~al.,}{{Ba{\~n}ados}
  et~al.}{2015}]{Banados15}
{Ba{\~n}ados} E.,  et~al., 2015, \mn@doi [ApJ] {10.1088/0004-637X/804/2/118},
  \href {http://cdsads.u-strasbg.fr/abs/2015ApJ...804..118B} {804, 118}

\bibitem[\protect\citeauthoryear{Baek, Di~Matteo, Semelin, Combes  \&
  Revaz}{Baek et~al.}{2009}]{Baek09}
Baek S.,  Di~Matteo P.,  Semelin B.,  Combes F.,   Revaz Y.,  2009, A\&A, 495,
  389

\bibitem[\protect\citeauthoryear{{Baek}, {Semelin}, {Di Matteo}, {Revaz}  \&
  {Combes}}{{Baek} et~al.}{2010}]{Baek10}
{Baek} S.,  {Semelin} B.,  {Di Matteo} P.,  {Revaz} Y.,   {Combes} F.,  2010,
  \mn@doi [A\&A] {10.1051/0004-6361/201014347}, \href
  {http://adsabs.harvard.edu/abs/2010A%26A...523A...4B} {523, A4+}

\bibitem[\protect\citeauthoryear{{Campisi}, {Maio}, {Salvaterra}  \&
  {Ciardi}}{{Campisi} et~al.}{2011}]{Campisi11}
{Campisi} M.~A.,  {Maio} U.,  {Salvaterra} R.,   {Ciardi} B.,  2011, \mn@doi
  [MNRAS] {10.1111/j.1365-2966.2011.19238.x}, \href
  {http://cdsads.u-strasbg.fr/abs/2011MNRAS.416.2760C} {416, 2760}

\bibitem[\protect\citeauthoryear{{Carilli}, {Gnedin}  \& {Owen}}{{Carilli}
  et~al.}{2002}]{Carilli02}
{Carilli} C.~L.,  {Gnedin} N.~Y.,   {Owen} F.,  2002, \mn@doi [ApJ]
  {10.1086/342179}, \href {http://cdsads.u-strasbg.fr/abs/2002ApJ...577...22C}
  {577, 22}

\bibitem[\protect\citeauthoryear{Chuzhoy \& Zheng}{Chuzhoy \&
  Zheng}{2007}]{Chuzhoy07}
Chuzhoy L.,  Zheng Z.,  2007, ApJ, 670, 912

\bibitem[\protect\citeauthoryear{{Ciardi} et~al.,}{{Ciardi}
  et~al.}{2013}]{Ciardi13}
{Ciardi} B.,  et~al., 2013, \mn@doi [MNRAS] {10.1093/mnras/sts156}, \href
  {http://adsabs.harvard.edu/abs/2013MNRAS.428.1755C} {428, 1755}

\bibitem[\protect\citeauthoryear{{Ciardi} et~al.,}{{Ciardi}
  et~al.}{2015}]{Ciardi15}
{Ciardi} B.,  et~al., 2015, preprint, \href
  {http://cdsads.u-strasbg.fr/abs/2015arXiv150407448C} {} (\mn@eprint {arXiv}
  {1504.07448})

\bibitem[\protect\citeauthoryear{{Ewall-Wice}, {Dillon}, {Mesinger}  \&
  {Hewitt}}{{Ewall-Wice} et~al.}{2014}]{Ewall-Wice14}
{Ewall-Wice} A.,  {Dillon} J.~S.,  {Mesinger} A.,   {Hewitt} J.,  2014, \mn@doi
  [MNRAS] {10.1093/mnras/stu666}, \href
  {http://cdsads.u-strasbg.fr/abs/2014MNRAS.441.2476E} {441, 2476}

\bibitem[\protect\citeauthoryear{{Fan} et~al.,}{{Fan} et~al.}{2006}]{Fan06}
{Fan} X.,  et~al., 2006, \mn@doi [AJ] {10.1086/504836}, \href
  {http://cdsads.u-strasbg.fr/abs/2006AJ....132..117F} {132, 117}

\bibitem[\protect\citeauthoryear{{Fialkov} \& {Barkana}}{{Fialkov} \&
  {Barkana}}{2014}]{Fialkov14b}
{Fialkov} A.,  {Barkana} R.,  2014, \mn@doi [MNRAS] {10.1093/mnras/stu1744},
  \href {http://cdsads.u-strasbg.fr/abs/2014MNRAS.445..213F} {445, 213}

\bibitem[\protect\citeauthoryear{Field}{Field}{1958}]{Field58}
Field G.,  1958, Proc. IRE, 46, 240

\bibitem[\protect\citeauthoryear{{Furlanetto}}{{Furlanetto}}{2006}]{Furlanetto06c}
{Furlanetto} S.~R.,  2006, \mn@doi [MNRAS] {10.1111/j.1365-2966.2006.10603.x},
  \href {http://cdsads.u-strasbg.fr/abs/2006MNRAS.370.1867F} {370, 1867}

\bibitem[\protect\citeauthoryear{{Furlanetto} \& {Loeb}}{{Furlanetto} \&
  {Loeb}}{2002}]{Furlanetto02}
{Furlanetto} S.~R.,  {Loeb} A.,  2002, \mn@doi [ApJ] {10.1086/342757}, \href
  {http://adsabs.harvard.edu/abs/2002ApJ...579....1F} {579, 1}

\bibitem[\protect\citeauthoryear{{Furlanetto} \& {Pritchard}}{{Furlanetto} \&
  {Pritchard}}{2006}]{Furlanetto06b}
{Furlanetto} S.~R.,  {Pritchard} J.~R.,  2006, \mn@doi [MNRAS]
  {10.1111/j.1365-2966.2006.10899.x}, \href
  {http://cdsads.u-strasbg.fr/abs/2006MNRAS.372.1093F} {372, 1093}

\bibitem[\protect\citeauthoryear{Furlanetto, Oh  \& Briggs}{Furlanetto
  et~al.}{2006}]{Furlanetto06}
Furlanetto S.~R.,  Oh S.~P.,   Briggs F.~H.,  2006, PhR, 433, 181

\bibitem[\protect\citeauthoryear{{Ghisellini}, {Haardt}, {Ciardi}, {Sbarrato},
  {Gallo}, {Tavecchio}  \& {Celotti}}{{Ghisellini} et~al.}{2015}]{Ghisellini15}
{Ghisellini} G.,  {Haardt} F.,  {Ciardi} B.,  {Sbarrato} T.,  {Gallo} E.,
  {Tavecchio} F.,   {Celotti} A.,  2015, preprint, \href
  {http://cdsads.u-strasbg.fr/abs/2015arXiv150505512G} {} (\mn@eprint {arXiv}
  {1505.05512})

\bibitem[\protect\citeauthoryear{{Gnedin}}{{Gnedin}}{2000}]{Gnedin00}
{Gnedin} N.~Y.,  2000, \mn@doi [ApJ] {10.1086/308876}, \href
  {http://cdsads.u-strasbg.fr/abs/2000ApJ...535..530G} {535, 530}

\bibitem[\protect\citeauthoryear{{Haiman}, {Quataert}  \& {Bower}}{{Haiman}
  et~al.}{2004}]{Haiman04}
{Haiman} Z.,  {Quataert} E.,   {Bower} G.~C.,  2004, \mn@doi [ApJ]
  {10.1086/422834}, \href {http://cdsads.u-strasbg.fr/abs/2004ApJ...612..698H}
  {612, 698}

\bibitem[\protect\citeauthoryear{Hirata}{Hirata}{2006}]{Hirata06}
Hirata C.~M.,  2006, MNRAS, 367, 259

\bibitem[\protect\citeauthoryear{Iliev, Mellema, Pen, Merz, Shapiro  \&
  Alvarez}{Iliev et~al.}{2006}]{Iliev06}
Iliev I.~T.,  Mellema G.,  Pen U.-L.,  Merz H.,  Shapiro P.~R.,   Alvarez
  M.~A.,  2006, MNRAS, 369, 1625

\bibitem[\protect\citeauthoryear{Iliev, Whalen, Mellema, Ahn  \& Baek}{Iliev
  et~al.}{2009}]{Iliev09}
Iliev I.~T.,  Whalen D.,  Mellema G.,  Ahn K.,   Baek S.,  2009, MNRAS, 400,
  1283

\bibitem[\protect\citeauthoryear{Kuhlen, Maudau  \& Montgomery}{Kuhlen
  et~al.}{2006}]{Kuhlen06}
Kuhlen M.,  Maudau P.,   Montgomery R.,  2006, ApJ, 637, 1

\bibitem[\protect\citeauthoryear{{Mack} \& {Wyithe}}{{Mack} \&
  {Wyithe}}{2012}]{Mack12}
{Mack} K.~J.,  {Wyithe} J.~S.~B.,  2012, \mn@doi [MNRAS]
  {10.1111/j.1365-2966.2012.21561.x}, \href
  {http://adsabs.harvard.edu/abs/2012MNRAS.425.2988M} {425, 2988}

\bibitem[\protect\citeauthoryear{{Madau}, {Meiksin}  \& {Rees}}{{Madau}
  et~al.}{1997}]{Madau97}
{Madau} P.,  {Meiksin} A.,   {Rees} M.~J.,  1997, ApJ, \href
  {http://adsabs.harvard.edu/abs/1997ApJ...475..429M} {475, 429}

\bibitem[\protect\citeauthoryear{{Mao}, {Shapiro}, {Mellema}, {Iliev}, {Koda}
  \& {Ahn}}{{Mao} et~al.}{2012}]{Mao12}
{Mao} Y.,  {Shapiro} P.~R.,  {Mellema} G.,  {Iliev} I.~T.,  {Koda} J.,   {Ahn}
  K.,  2012, \mn@doi [MNRAS] {10.1111/j.1365-2966.2012.20471.x}, \href
  {http://cdsads.u-strasbg.fr/abs/2012MNRAS.422..926M} {422, 926}

\bibitem[\protect\citeauthoryear{{Meiksin}}{{Meiksin}}{2011}]{Meiksin11}
{Meiksin} A.,  2011, \mn@doi [MNRAS] {10.1111/j.1365-2966.2011.19362.x}, \href
  {http://adsabs.harvard.edu/abs/2011MNRAS.417.1480M} {417, 1480}

\bibitem[\protect\citeauthoryear{Mellema, Iliev, Alvarez  \& Shapiro}{Mellema
  et~al.}{2006}]{Mellema06}
Mellema G.,  Iliev I.~T.,  Alvarez M.,   Shapiro P.~R.,  2006, NewA, 11, 374

\bibitem[\protect\citeauthoryear{{Mellema} et~al.,}{{Mellema}
  et~al.}{2013}]{Mellema13}
{Mellema} G.,  et~al., 2013, \mn@doi [ExA] {10.1007/s10686-013-9334-5}, \href
  {http://cdsads.u-strasbg.fr/abs/2013ExA...tmp...17M} {}

\bibitem[\protect\citeauthoryear{{Mesinger}}{{Mesinger}}{2010}]{Mesinger10}
{Mesinger} A.,  2010, \mn@doi [MNRAS] {10.1111/j.1365-2966.2010.16995.x}, \href
  {http://adsabs.harvard.edu/abs/2010MNRAS.407.1328M} {407, 1328}

\bibitem[\protect\citeauthoryear{{Mesinger}, {Furlanetto}  \& {Cen}}{{Mesinger}
  et~al.}{2011}]{Mesinger11}
{Mesinger} A.,  {Furlanetto} S.,   {Cen} R.,  2011, \mn@doi [MNRAS]
  {10.1111/j.1365-2966.2010.17731.x}, \href
  {http://adsabs.harvard.edu/abs/2011MNRAS.411..955M} {411, 955}

\bibitem[\protect\citeauthoryear{{Planck Collaboration} et~al.,}{{Planck
  Collaboration} et~al.}{2015}]{Planck15}
{Planck Collaboration} et~al., 2015, preprint, \href
  {http://adsabs.harvard.edu/abs/2015arXiv150201589P} {} (\mn@eprint {arXiv}
  {1502.01589})

\bibitem[\protect\citeauthoryear{{Santos}, {Ferramacho}, {Silva}, {Amblard}  \&
  {Cooray}}{{Santos} et~al.}{2010}]{Santos10}
{Santos} M.~G.,  {Ferramacho} L.,  {Silva} M.~B.,  {Amblard} A.,   {Cooray} A.,
   2010, \mn@doi [MNRAS] {10.1111/j.1365-2966.2010.16898.x}, \href
  {http://adsabs.harvard.edu/abs/2010MNRAS.406.2421S} {406, 2421}

\bibitem[\protect\citeauthoryear{Semelin \& Combes}{Semelin \&
  Combes}{2002}]{Semelin02}
Semelin B.,  Combes F.,  2002, A\&A, 495, 389

\bibitem[\protect\citeauthoryear{Semelin, Combes  \& Baek}{Semelin
  et~al.}{2007}]{Semelin07}
Semelin B.,  Combes F.,   Baek S.,  2007, A\&A, 495, 389

\bibitem[\protect\citeauthoryear{{Shimabukuro}, {Ichiki}, {Inoue}  \&
  {Yokoyama}}{{Shimabukuro} et~al.}{2014}]{Shimabukuro14}
{Shimabukuro} H.,  {Ichiki} K.,  {Inoue} S.,   {Yokoyama} S.,  2014, \mn@doi
  [Phys. Rev. D] {10.1103/PhysRevD.90.083003}, \href
  {http://cdsads.u-strasbg.fr/abs/2014PhRvD..90h3003S} {90, 083003}

\bibitem[\protect\citeauthoryear{{Vasiliev} \& {Shchekinov}}{{Vasiliev} \&
  {Shchekinov}}{2013}]{Vasiliev13}
{Vasiliev} E.~O.,  {Shchekinov} Y.~A.,  2013, \mn@doi [ApJ]
  {10.1088/0004-637X/777/1/8}, \href
  {http://cdsads.u-strasbg.fr/abs/2013ApJ...777....8V} {777, 8}

\bibitem[\protect\citeauthoryear{{Vonlanthen}, {Semelin}, {Baek}  \&
  {Revaz}}{{Vonlanthen} et~al.}{2011}]{Vonlanthen11}
{Vonlanthen} P.,  {Semelin} B.,  {Baek} S.,   {Revaz} Y.,  2011, \mn@doi [A\&A]
  {10.1051/0004-6361/201116811}, \href
  {http://adsabs.harvard.edu/abs/2011A%26A...532A..97V} {532, A97+}

\bibitem[\protect\citeauthoryear{{Wilman} et~al.,}{{Wilman}
  et~al.}{2008}]{Wilman08}
{Wilman} R.~J.,  et~al., 2008, \mn@doi [MNRAS]
  {10.1111/j.1365-2966.2008.13486.x}, \href
  {http://cdsads.u-strasbg.fr/abs/2008MNRAS.388.1335W} {388, 1335}

\bibitem[\protect\citeauthoryear{Wouthuysen}{Wouthuysen}{1952}]{Wouthuysen52}
Wouthuysen S.~A.,  1952, AJ, 57, 21

\bibitem[\protect\citeauthoryear{{Xu}, {Chen}, {Fan}, {Trac}  \& {Cen}}{{Xu}
  et~al.}{2009}]{Xu09}
{Xu} Y.,  {Chen} X.,  {Fan} Z.,  {Trac} H.,   {Cen} R.,  2009, \mn@doi [ApJ]
  {10.1088/0004-637X/704/2/1396}, \href
  {http://cdsads.u-strasbg.fr/abs/2009ApJ...704.1396X} {704, 1396}

\bibitem[\protect\citeauthoryear{{Xu}, {Ferrara}  \& {Chen}}{{Xu}
  et~al.}{2011}]{Xu11}
{Xu} Y.,  {Ferrara} A.,   {Chen} X.,  2011, \mn@doi [MNRAS]
  {10.1111/j.1365-2966.2010.17579.x}, \href
  {http://adsabs.harvard.edu/abs/2011MNRAS.410.2025X} {410, 2025}

\bibitem[\protect\citeauthoryear{{Zawada}, {Semelin}, {Vonlanthen}, {Baek}  \&
  {Revaz}}{{Zawada} et~al.}{2014}]{Zawada14}
{Zawada} K.,  {Semelin} B.,  {Vonlanthen} P.,  {Baek} S.,   {Revaz} Y.,  2014,
  \mn@doi [MNRAS] {10.1093/mnras/stu035}, \href
  {http://cdsads.u-strasbg.fr/abs/2014MNRAS.439.1615Z} {439, 1615}

\makeatother
\end{thebibliography}

\appendix

\end{document}